# PCMSolver: an Open-Source Library for Solvation Modeling

Roberto Di Remigio[1*]   |   Arnfinn Hykkerud Steindal[1]   |
Krzysztof Mozgawa[1]   |   Ville Weijo[1]   |   Hui Cao[2]   |
Luca Frediani[1]

[1]Hylleraas Centre for Quantum Molecular Sciences, Department of Chemistry, University of Tromsø - The Arctic University of Norway, N-9037 Tromsø, Norway

[2]Jiangsu Key Lab. of Atmosph. Environment Monitoring and Pollution Control, Collaborative Center of Atmosph. Environment and Equipment Technology, School of Env. Sci. and Eng., Nanjing Univ. of Information Science and Technology, Nanjing 210044, P.R. China

**Correspondence**
Roberto Di Remigio PhD, Hylleraas Centre for Quantum Molecular Sciences, Department of Chemistry, University of Tromsø - The Arctic University of Norway, N-9037 Tromsø, Norway
Email: roberto.d.remigio@uit.no

**Present address**
*Department of Chemistry, Virginia Tech, Blacksburg, Virginia 24061, United States

**Funding information**
This work was partially supported by the Research Council of Norway through its Centres of Excellence scheme, project number 262695, and through its Mobility Grant scheme, project number 261873.

PCMSOLVER is an open-source library for continuum electrostatic solvation. It can be combined with any quantum chemistry code and requires a minimal interface with the host program, greatly reducing programming effort. As input, PCMSOLVER needs only the molecular geometry to generate the cavity and the expectation value of the molecular electrostatic potential on the cavity surface. It then returns the solvent polarization back to the host program. The design is powerful and versatile: minimal loss of performance is expected, and a standard single point self-consistent field implementation requires no more than 2 days of work. We provide a brief theoretical overview, followed by two tutorials: one aimed at quantum chemistry program developers wanting to interface their code with PCMSOLVER, the other aimed at contributors to the library. We finally illustrate past and ongoing work, showing the library's features, combined with several quantum chemistry programs.

**Abbreviations:** API, application programming interface; CI, continuous integration; PCM, polarizable continuum model; PR, pull request; DVCS, Distributed version control system

# 1 | INTRODUCTION

The past ten years have seen theoretical and computational methods become an invaluable complement to experiment in the practice of chemistry. Understanding experimental observations of chemical phenomena, ranging from reaction barriers to spectroscopies, requires proper *in silico* simulations to achieve insight into the fundamental processes at work. Quantum chemistry program packages have evolved to tackle this ever-increasing range of possible applications, with a particular focus on computational performance and scalability. These latter concerns have driven a large body of recent developments, but it has become apparent that similar efforts have to be devoted into the software development *infrastructure* and *practices*. Code bases in quantum chemistry have grown over a number of years, in most cases without an overarching vision on the architecture and design of the code. As more features continue to be added, the friction with legacy code bases makes itself felt: either the code undergoes a time-consuming rewrite or it becomes the domain of few experts. Both approaches are wasteful of resources and can seriously hinder the reproducibility of computational results. It is essential to find more effective ways of organizing scientific code and programming efforts in quantum chemistry. To be able to manage the growing complexity of quantum chemical program packages, the keywords *efficiency* and *scalability* have to be compounded with *maintainability* and *extensibility*. The sustainability of software development in the computational sciences has become a reason for growing concern, especially because reproducibility of results could suffer [60, 59, 61, 71, 92, 110, 70, 144, 62, 136, 145, 11, 9, 10, 7, 14].

The paradigm of *modular programming* has been one of the emerging motifs in modern scientific software development. In computer science, the idea is not new. Dijkstra and Parnas advocated it as early as 1968 in the development of operating systems [40, 99]. Dividing a complex system into smaller, more manageable portions is a very effective strategy. It reduces the overall complexity, cognitive load and ultimately the likelihood of introducing faults into software. Sets of functionalities are isolated into libraries, with well-defined application programmers interfaces (APIs). The implementation of clearly defined computational tasks into separate, independent pieces of software guarantees that the development of conceptually different functionalities does not get inextricably and unnecessarily entangled. Each library becomes a computational black box that is *developed*, *tested*, *packaged* and *distributed* independently from any of the programs that might potentially use it. The BLAS and LAPACK sets of subroutines for linear algebra are certainly success stories for the modular approach to software development. Well-crafted APIs are key to delimiting the problem domain. Eventually, as happened for BLAS and LAPACK, they enforce a standardization of the functionality offered [116], such that one implementation can be interchanged for another without the need to rewrite any code.

The polarizable continuum model (PCM) is a continuum solvation model introduced in quantum chemistry (QC) in the 80s [94] and actively developed ever since [140, 91]. Its simple formulation and ease of implementation have made it the go-to method when a quick estimate of solvation effects is desired. The clear separation between the solvation and the quantum chemical layers of a calculation, make it an ideal candidate for the design and implementation of an API for classical polarizable solvation models. The input to and output from such a library are clear and well-defined affording a natural API design that can straightforwardly be compared with the working equations of the method.

We here present the open-source PCMSOLVER library, which we have developed over the past few years conforming to the principles just outlined. With PCMSOLVER, we aim at providing the QC development community with a reliable and easy-to-use implementation of the PCM. The library is released under the terms of the version 3 of the GNU Lesser General Public Licence (LGPL) [137], to guarantee a lower threshold to adoption and to encourage third-party contributions. Our design choices allow for the fast development of interfaces with *any* existing QC code with negligible coding effort and run-time performance penalty. In order to describe the implementation of PCMSOLVER, we will recap its theoretical foundations in section 2. We are not aiming at a detailed exposition, but we will rather emphasize the aspects which are important in connection with the development of an independent library for solvation. We will show how the PCM provides a unified blueprint for all classical polarizable models by making use of the variational formulation introduced by Lipparini et al. [86]. Section 3 will offer an high-level overview of the library and a step-by-step tutorial for QC program developers on how to interface with PCMSOLVER. Section 4 will dive deeper into the internal structure of the library, discuss the various components and their interaction. This detailed tutorial is aimed at potential contributors to the library and is complemented by section 5, discussing the licensing model and the contribution workflow. In section 6, we will present a few applications of PCMSOLVER, drawing on past and ongoing work in our group using different QC program packages. Section 7 will present a summary and an overview of the work ahead.

## 2 | THEORY

The original idea of the PCM is to describe solute-solvent interactions only by means of electrostatics and polarization between the solute molecule and the solvent. The solvent is modeled as a dielectric continuum with a given permittivity $\varepsilon$. A cavity $\Omega_i$, with closed boundary $\Gamma \equiv \partial\Omega_i$, is built inside this medium and the solute is placed in it (see figure 1). Quantum mechanics is used to describe the solute. Within the Born-Oppenheimer approximation, the nuclei are kept fixed, whereas the electrons are described by either density-functional theory (DFT) or wave function theory (WFT). For a given electronic density and fixed nuclear positions, the vacuum molecular electrostatic potential (MEP) $\varphi(r)$ is fully determined for all points $r$ in space. The interaction between the molecule and the solute becomes a problem of classical electrostatics: the source density $\rho(r)$ and the dielectric continuum mutually polarize. The generalized Pois-son equation for a medium with a position-dependent permittivity $\varepsilon(r)$ is the governing equation for this transmission problem [123]

$$\nabla \cdot [\varepsilon(r) \nabla u(r)] = -4\pi\rho(r) = -4\pi \left( \sum_{A=1}^{N_{\text{nuclei}}} Z_A \delta(r - R_A) - \rho_e(r) \right) \qquad (1)$$

where $u(\mathbf{r})$ is now the electrostatic potential in space including the polarization of the continuum. The information about the medium and the cavity is all encoded in the dielectric permittivity function $\varepsilon(\mathbf{r})$, which is equal to 1 inside the cavity and depends on the medium outside. The generalized Poisson equation admits a unique solution, once the boundary conditions at the cavity and at infinity are fixed [32]. In the simplest case of a uniform, isotropic and homogeneous dielectric outside the cavity with permittivity $\varepsilon$ (scalar and position-independent), the problem simplifies to the solution of the following set of equations:

$$\nabla^2 u(\mathbf{r}) = -4\pi\rho(\mathbf{r}) \quad \forall \mathbf{r} \in C \tag{2a}$$

$$\varepsilon\nabla^2 u(\mathbf{r}) = 0 \quad \forall \mathbf{r} \notin C \tag{2b}$$

$$\lim_{|\mathbf{r}|\to\Gamma^+} u(\mathbf{r}) = \lim_{|\mathbf{r}|\to\Gamma^-} u(\mathbf{r}) \tag{2c}$$

$$\varepsilon \lim_{|\mathbf{r}|\to\Gamma^+} \frac{\partial u(\mathbf{r})}{\partial \mathbf{n}} = \lim_{|\mathbf{r}|\to\Gamma^-} \frac{\partial u(\mathbf{r})}{\partial \mathbf{n}} \tag{2d}$$

$$|u| \le C\|\mathbf{x}\|^{-1} \text{ for } \|\mathbf{x}\| \to \infty \tag{2e}$$

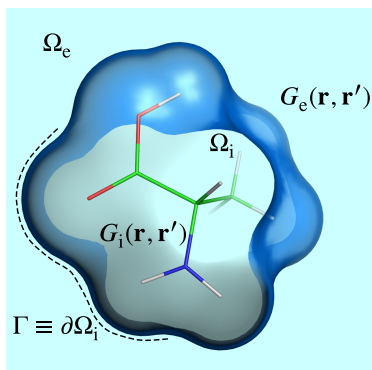

**FIGURE 1** The physical setting of the PCM. The molecular solute is represented by its charge density $\rho_i$ assumed to be fully enclosed in a cavity $\Omega_i$ with boundary $\Gamma \equiv \partial\Omega_i$. The permittivity inside the cavity is that of vacuum, $\varepsilon = 1$, and hence has Green's function $G_i = \frac{1}{|r-r'|}$. The cavity is carved out of an infinite, structureless continuum with Green's function $G_e$ determined by its material properties. The exterior volume $\Omega_e$ is completely filled by the continuum.

The first two equations are a simple rewrite of the original Poisson equation inside and outside the cavity, respectively. Equations (2c) and (2d) are the boundary conditions for the electrostatic potential and its normal derivative (electrostatic field) at the cavity boundary. The last equation is the radiation condition at infinity. It is beyond the scope of this contribution to discuss the general solution strategy of such a problem in depth and we refer the reader to the abundant literature on the subject [32, 69, 123]. In the integral equation formalism (IEF), we express the mutual solute-solvent polarization in terms of an apparent surface charge (ASC) $\sigma(\mathbf{s})$ for all points $\mathbf{s}$ on the cavity surface $\Gamma \equiv \partial\Omega_\mathrm{i}$, that is, the ASC is *entirely* supported on the cavity boundary $\Gamma$ achieving a reduction in the dimensionality of the electrostatic problem to be solved. The set of equations (2) is then reformulated as an integral equation on the cavity boundary:

$$\hat{\mathcal{T}}\sigma(\mathbf{s}) = -\hat{\mathcal{R}}\varphi(\mathbf{s}). \tag{3}$$

For a uniform, isotropic and homogeneous dielectric, the $\hat{\mathcal{T}}$ and $\hat{\mathcal{R}}$ boundary integral (BI) operators are defined as:

$$\hat{\mathcal{T}} = \left(2\pi\frac{\varepsilon+1}{\varepsilon-1}\hat{\mathcal{I}} - \hat{\mathcal{D}}\right)\hat{\mathcal{S}} \tag{4a}$$

$$\hat{\mathcal{R}} = \left(2\pi\hat{\mathcal{I}} - \hat{\mathcal{D}}\right) \tag{4b}$$

where $\hat{\mathcal{I}}$ is the identity operator and $\hat{\mathcal{S}}$, $\hat{\mathcal{D}}$ are components of the Calderón projector. Such operators are completely defined once the cavity geometry and the dielectric properties of the medium are known and form the cornerstone of any implementation of IEF-PCM. Three of the four components of the projector are needed for the IEF-PCM [25, 69, 123]:

$$\left(\hat{\mathcal{S}}_\star u\right)(\mathbf{s}) = \int_\Gamma G_\star(\mathbf{s},\mathbf{s}')u(\mathbf{s}')\mathrm{d}\mathbf{s}' \tag{5a}$$

$$\left(\hat{\mathcal{D}}_\star u\right)(\mathbf{s}) = \int_\Gamma \varepsilon_\star(\mathbf{s}')\frac{\partial G_\star(\mathbf{s},\mathbf{s}')}{\partial n_{\mathbf{s}'}}u(\mathbf{s}')\mathrm{d}\mathbf{s}' \tag{5b}$$

$$\left(\hat{\mathcal{D}}_\star^\dagger u\right)(\mathbf{s}) = \int_\Gamma \varepsilon_\star(\mathbf{s})\frac{\partial G_\star(\mathbf{s},\mathbf{s}')}{\partial n_{\mathbf{s}}}u(\mathbf{s}')\mathrm{d}\mathbf{s}', \tag{5c}$$

the derivatives are taken in the direction of the outgoing normal vector to the point. The ⋆ index exemplifies that the internal or external Green's function can be used. The form of such operators is only dependent on the geometry of the molecular cavity and on the Green's function of the problem. Thanks to the IEF, the approach is not limited to uniform, isotropic and homogeneous dielectrics; any solvent for which it is possible to obtain a Green's function for the electrostatic problem is amenable to this treatment. Several media in addition to uniform dielectrics admit a Green's function in closed form: anisotropic dielectric (tensorial permittivity), ionic solutions (constant permittivity and ionic strength) [25], sharp planar [49] and spherical interfaces [31] (two permittivities). The Green's function for diffuse interfaces, where a smooth position-dependent permittivity function is used, can be built numerically[48, 37]. Most of these environments are provided by PCMSOLVER and the missing ones are under development. Table 1 gives a compact overview. The Green's function component of PCMSOLVER is designed to handle functions that can be expressed as the sum of a singular and a nonsingular component:

$$G(\mathbf{r}, \mathbf{r}') = \mathcal{F}_\varepsilon(\mathbf{r}, \mathbf{r}') + G_{\text{img}}(\mathbf{r}, \mathbf{r}'). \tag{6}$$

The first $\mathcal{F}_\varepsilon(\mathbf{r}, \mathbf{r}')$ presents a Coulomb singularity, possibly modulated by an effective permittivity – $\beta(\mathbf{r}, \mathbf{r}')$ – which depends on the positions of the source $\mathbf{r}$ and the probe $\mathbf{r}'$:

$$\mathcal{F}_\varepsilon(\mathbf{r}, \mathbf{r}') \simeq \beta(\mathbf{r}, \mathbf{r}') \frac{1}{|\mathbf{r} - \mathbf{r}'|} \tag{7}$$

The second, nonsingular component, when present, is generically referred to as the *image*. In some cases it can be written in a closed form (*e.g.* sharp interfaces), whereas in others (*e.g.* diffuse interfaces) a numerical integration of an ordinary differential equation (ODE) is required.

**TABLE 1** Green's functions for different dielectric media and their availability within PCMSOLVER.

| Medium | Parameters | Differential equation | Green's function | Notes |
|---|---|---|---|---|
| Uniform dielectric | $\varepsilon$ | $-\varepsilon\nabla^2 V(\mathbf{r}) = 0$ | $\frac{1}{\varepsilon|\mathbf{r}-\mathbf{r}'|}$ | |
| Ionic solution | $\varepsilon, \kappa$ | $-\varepsilon(\nabla^2 - \kappa^2)V(\mathbf{r}) = 0$ | $\frac{e^{-\kappa|\mathbf{r}-\mathbf{r}'|}}{\varepsilon|\mathbf{r}-\mathbf{r}'|}$ | Linearized Poisson-Boltzmann equation, valid in the regime of small ionic strenghts. |
| Anisotropic dielectric | $(\underline{\underline{\varepsilon}})_{ij}, i,j = 1, 2, 3$ | $-\nabla \cdot (\underline{\underline{\varepsilon}}\nabla V(\mathbf{r})) = 0$ | $\frac{1}{\sqrt{\det \underline{\underline{\varepsilon}}(\mathbf{r}\cdot\underline{\underline{\varepsilon}}^{-1}\mathbf{r})}}$ | A tensorial permittivity is a model applicable to liquid crystals |
| Sharp planar interface | $\varepsilon_1(z<0)$ and $\varepsilon_2(z>0)$ | $-\varepsilon_i\nabla^2 V(\mathbf{r}) = 0$ | $\frac{1}{\varepsilon_1|\mathbf{r}-\mathbf{r}'|} + \frac{\varepsilon_1-\varepsilon_2}{\varepsilon_1+\varepsilon_2}\frac{1}{|\mathbf{r}-\mathbf{r}'|}$ | The reported expression is valid for source and probe located in medium 1. See for instance Ref.[72] for the other cases. |
| Sharp spherical interface[93, 31] | $\varepsilon_1(r<r_0)$ and $\varepsilon_2(r>r_0)$ | $-\varepsilon_i\nabla^2 V(\mathbf{r}) = 0$ | $\frac{1}{\varepsilon_2|\mathbf{r}-\mathbf{r}'|}$ $+ \frac{1}{\varepsilon_2}\left\{\sum_{\ell=1}^{\infty}\frac{a^{2\ell+1}}{b^{\ell+1}}C_\ell\frac{P_\ell(\cos\gamma)}{|\mathbf{r}'-\mathbf{r}_s|^{\ell+1}}\right\}$ | The reported expression is valid for source and probe located in medium 2 (outside the sphere). The other cases have not yet been considered. |
| Diffuse planar interface | $\varepsilon(z)$ | $-\nabla\cdot\varepsilon(z)\nabla V = 0$ | $\frac{1}{C(z,z')|\mathbf{r}-\mathbf{r}'|} + G_{im}(\mathbf{r},\mathbf{r}')$ | Effective dielectric constant $C(z,z')$ and image potential obtained by numerical integration (cylindrical coordinates), followed by convolution with Bessel function $J_0$ [48]. |
| Diffuse spherical interface | $\varepsilon(r)$ | $-\nabla\cdot\varepsilon(r)\nabla V = 0$ | $\frac{1}{C(r,r')|\mathbf{r}-\mathbf{r}'|} + G_{im}(\mathbf{r},\mathbf{r}')$ | Effective dielectric constant $C(r,r')$ and image potential obtained by numerical integration in spherical coordinates, followed by a summation in spherical harmonics [37]. |

## 2.1 | The boundary element method

The practical solution of Eq. (3) is achieved by means of the boundary element method (BEM). The cavity boundary is discretized into $N_{\text{mesh}}$ finite elements – $T_i$ – by a meshing algorithm that generates polygonal finite elements. Triangles or quadrangles are the most usual choices and the finite elements can be either planar or curved. The mathematical framework for the BEM is provided by Galerkin approximation theory [56, 43, 123]. The application of any integral operator $\hat{\mathcal{A}}$ with kernel $k_A(\mathbf{s}, \mathbf{s}')$ on a function $f(\mathbf{s})$ supported on the boundary:

$$(\hat{\mathcal{A}}f)(\mathbf{s}) = \int_\Gamma d\mathbf{s}' k_A(\mathbf{s}, \mathbf{s}') f(\mathbf{s}') \tag{8}$$

can be discretized as:

$$A_{ij} f_j = \int_{T_i} d\mathbf{s} \int_{T_j} d\mathbf{s}' k_A(\mathbf{s}, \mathbf{s}') f(\mathbf{s}'). \tag{9}$$

The choice of the basis functions on the mesh and of the integration procedure will determine the properties of the BEM adopted, including its accuracy. Note that if singular kernels arise in the theory, proper care will have to be taken in calculating matrix elements for close or identical pairs of finite elements $T_i$, $T_j$. Thus, discretization of the surface induces a discretization of the operators involved in the IEF equation (3). The integral operators are represented as matrices, whereas the functions supported on the cavity boundary become vectors: the problem is recast as a system of linear equations.

The current version of PCMSOLVER implements a straightforward centroid collocation method: for each finite element $i$, the charge density $\sigma$ is condensed in a point charge $q_i$. The off-diagonal matrix elements of the Calderón projector components are then simply obtained as the value of the Green's function and its derivatives at those points. For instance $\hat{\mathcal{S}}_{\star,ij} = G_\star(\mathbf{s}_i, \mathbf{s}_j)$. Because of the divergence in the kernels, it is clear that such a discretization will break down if naïvely applied in the calculation of the diagonal elements. These singularities are however integrable and thus methods have been formulated to overcome this difficulty. In the traditional PCM implementation, the analytic form available for a polar cap is fitted and parametrized to a polygonal patch [94, 140]. For the $\hat{\mathcal{D}}$ operator, sum rules, relating the diagonal elements to their respective row or column, have been derived by Purisima and Nilar [112, 111]. For Green's functions not available in closed-form, such as the diffuse interfaces, particular care needs to be taken to isolate the singularity. The partition in equation (6) proves particularly convenient. The singularity, known in closed-form, is then taken care of by one of the methods above, whereas the nonsingular remainder is integrated by standard quadrature methods. Gaussian quadrature for the centroid collocation of the diagonal elements has also been discussed in the literature [25]. The more sophisticated wavelet Galerkin method uses numerical quadrature for the calculation of all matrix elements [143, 23]. The singularities are treated using the Duffy trick [123, 117] instead of parametrized approximate formulas.

## 2.2 | Variational formulation of the PCM

The introduction of the variational formulation is also an important recent development for the PCM formalism. Lipparini et al. have shown that it is possible to express the polarization problem of the IEF as the minimization of the appropriate functional [86]. This reformulation is possible for any elliptic partial differential equation [43]. For exam-ple, the minimum of the functional:

$$\mathcal{F}(\varphi) = \frac{1}{2} \langle \nabla \varphi | \epsilon(r) | \nabla \varphi \rangle - 4\pi \langle \varphi | \rho \rangle \tag{10}$$

corresponds to the unique solution of the generalized Poisson equation (1). For a general, position-dependent permittivity function the solution can be obtained as described by Fosso-Tande and Harrison [47]. It is also possible to recast the corresponding boundary integral equation into a variational problem, given the appropriate functional and functional spaces. Lipparini et al. [86] proposed the functional:

$$\mathcal{G}(\sigma) = \frac{1}{2} \left( \sigma, \hat{\mathcal{R}}^{-1}\hat{\mathcal{T}}\sigma \right)_\Gamma + (\varphi, \sigma)_\Gamma , \tag{11}$$

and proved that its minimum corresponds to the solution of the IEF-PCM equation (3). Here $(\cdot, \cdot)_\Gamma$ is the inner product in the suitable Sobolev space with support on the cavity boundary $\Gamma$ [69]. A variational formulation has several formal and practical advantages [20, 75, 132, 74, 85]:

1. It removes the non-linear coupling with the quantum mechanics (QM) problem, since the polarization charge density is optimized on the same footing as the QM parameters, *e.g.* orbitals in self-consistent field (SCF) theories.
2. It provides a unified framework to include continuum solvation regardless of the method used (molecular mechanics (MM), QM or both) simplifying the description of the coupling.
3. It simplifies the framework for the calculation of molecular properties.
4. It is convenient to include solvation in an extended Lagrangian formulation for molecular dynamics (MD) simulations.
5. It can be employed for other kinds of solvation methods (*e.g.* polarizable MM) with minimal modifications.

Both response theory for molecular properties and coupled cluster (CC) for correlated calculations, can be formulated using a Lagrangian formalism [129, 65]. In response theory, the quasienergy formalism [98, 27, 64] is employed to obtain linear and nonlinear molecular properties as high-order derivatives of a quasienergy Lagrangian. Such a Lagrangian can be formulated in the molecular orbital (MO) or atomic orbital (AO) basis, the latter allowing for an open-ended, recursive formulation and implementation of SCF-level molecular properties [139, 120]. In the variational formulation, the PCM ASC are just an additional variational parameter, on the same footing as the AO density matrix and the derivation of the response equations and properties expression to any order becomes a straightforward extension of the vacuum case [35]. As an example, the quadratic response function can be written as:

$$\begin{aligned}\langle\langle A; B, C \rangle\rangle_{\omega_b, \omega_c} = \frac{d\{\tilde{L}^a(\tilde{D}, \tilde{\sigma}, t)\}_T}{d\epsilon_b d\epsilon_c} &= L^{abc} \stackrel{\{Tr\}_T}{=} \mathcal{G}^{00,abc} + \mathcal{G}^{10,ac}\mathbf{D}^b + \mathcal{G}^{10,ab}\mathbf{D}^c \\ &+ \mathcal{G}^{20,a}\mathbf{D}^b\mathbf{D}^c + \mathcal{G}^{10,a}\mathbf{D}^{bc} + \mathcal{G}^{11,a}\mathbf{D}^b\sigma^c \\ &+ \mathcal{G}^{01,ac}\sigma^b + \mathcal{G}^{01,ab}\sigma^c + \mathcal{G}^{02,a}\sigma^b\sigma^c + \mathcal{G}^{01,a}\sigma^{bc} + \mathcal{G}^{11,a}\sigma^b\mathbf{D}^c \\ &- S^{abc}W - S^{ab}W^c - S^{ac}W^b - S^a W^{bc}\end{aligned} \tag{12}$$

where $\mathcal{G}$ is the solvation free energy functional, $\mathbf{D}$ is the density matrix, $\mathbf{S}$ is the overlap matrix, $\mathbf{W}$ is the energy-weighted density matrix, $\sigma$ is the ASC. In our notation, the indices $a, b, c$ represent derivatives with respect to the external perturbations, whereas the numerical indices 0, 1, 2 are derivatives with respect to the density matrix (first index) and the ASC (second index). For details about the derivation of the expression above we refer to the original manuscript [35]. We highlight here the symmetry in **D** and σ in the expression for the property, which greatly simplifies the derivation of the response equations and their subsequent implementation.

In combination with a CC wave function, the variational formalism is a powerful tool to derive the working equations of the method and identify more efficient approximations. More in detail, the formulation of a consistent many-body perturbation theory (MBPT) including solvent effects from a classical polarizable medium is simplified. Since the polarization does no longer depend nonlinearly on the CC density, it is much easier to identify at which perturbative order in the fluctuation potential the different PCM contributions play a role [33, 36]. The effective PCM-CC Lagrangian is the sum of the regular CC Lagrangian and the polarization energy functional [24, 26]:

$$\mathcal{L}_{\text{eff}}(t, \bar{t}, \sigma)_{\mathcal{M}} = \langle \text{HF}|e^{-T}H_0 e^T|\text{HF}\rangle + \sum_{u=1}^{\mathcal{M}} \langle \bar{t}_u|e^{-T}H_0 e^T|\text{HF}\rangle \\ + \frac{1}{2}\left(\sigma, \hat{\mathcal{R}}^{-1}\hat{\mathcal{T}}\sigma\right)_\Gamma + \left(\varphi_\text{N}(t,\bar{t})_\mathcal{M}, \sigma\right)_\Gamma + \left(\varphi_\text{N}(t,\bar{t})_\mathcal{M}, \sigma_\text{HF}\right)_\Gamma + U_\text{pol}^\text{ref}, \quad (13)$$

where $\mathcal{M}$ is the CC truncation level, $T$ the cluster operator and $\bar{t}$ the Lagrangian multipliers. Normal ordering [129], induces a natural separation between reference and correlation components of the MEP and ASC. The CC equations can then be obtained by differentiating the Lagrangian with respect to the variational parameters: $t$, $\bar{t}$ and σ. Note that the amplitudes and multipliers are now coupled through the MEP $\varphi_\text{N}(t,\bar{t})_\mathcal{M}$. Equation (13) is also the starting point for the formulation of CC perturbation theory (PT). Indeed we have shown that perturbative corrections for triple excitations for the PCM-CCSD can be easily derived in this framework [33, 36].

Several classical polarizable models besides the PCM introduce mutual solute-solvent polarization by means of a linear reaction field, leading to an energy functional of the form of Eq. (10). In particular, polarizable MM models are amenable to such a treatment [89]. The easiest alternative is constituted by the fluctuating charge (FQ) model which employs the same ingredients as PCM: the MEP and a set of fluctuating charges [119]. The expression for the energy functional is [82, 83]:

$$\mathcal{E}_\text{FQ} = \frac{1}{2}\boldsymbol{q} \cdot \boldsymbol{J} \cdot \boldsymbol{q} + \boldsymbol{q} \cdot \boldsymbol{\chi} + \boldsymbol{q} \cdot \boldsymbol{\lambda} + \boldsymbol{q} \cdot \boldsymbol{\varphi} \quad (14)$$

where the pairwise Coulomb repulsion matrix **J** and the electronegativity vector χ were introduced. Minimization of $\mathcal{E}_\text{FQ}$ yields the fluctuating charges **q**. Compared to the PCM functional in Eq. (11), $\mathcal{E}_\text{FQ}$ also contains the electronegativity parameters χ, describing the interaction of the charges with the other MM fragments and the Lagrange multipliers λ to ensure the electroneutrality of each separate fragment. The dependence on the external QM potential φ is otherwise identical, opening the way for an easy implementation of the model in PCMSOLVER with no modifications foreseen for the host program. As pointed out by Lipparini et al. [84], in a variational formalism layering different models becomes also straightforward: it will suffice to add the respective functionals and the interaction terms between each of them. For an FQ/PCM model this term is the electrostatic energy between the charges **q** and the surface polarization σ. The other widespread polarizable MM model makes use of fixed point multipoles and fluctuating dipoles at atomic sites [128]. The induced dipoles responding to the surrounding electrostatic field are the variational parameters. *Mutatis mutandis* it is possible to obtain a corresponding energy functional, although its implementation in PCMSOLVER and coupling with continuum solvation would require additional effort, in particular regarding the handling of the force field parameters and the polarization, since both are matrix, rather than vector, quantities.

## 2.3 | Coupling the classical and quantum problems

The coupling of the PCM with a SCF procedure can be achieved with the following step-by-step control flow for the final program:

1. The MEP $\varphi$ at the cavity points is computed by the host code:

$$\varphi(s_i) = \sum_{A=1}^{N_{\text{nuclei}}} \frac{Z_A}{|R_A - s_i|} + \sum_{\kappa\lambda} D_{\lambda\kappa} \int d\mathbf{r} \frac{-\Omega_{\kappa\lambda}(\mathbf{r})}{|\mathbf{r} - s_i|}, \quad \forall i = 1, N_{\text{mesh}} \qquad (15)$$

   where $D_{\lambda\kappa}$ and $\Omega_{\kappa\lambda}(\mathbf{r}) = \chi_\kappa^*(\mathbf{r})\chi_\lambda(\mathbf{r})$ are, respectively, the AO density and overlap distribution matrices. The MEP is passed to the PCM library.
2. The PCM library computes the ASC $\sigma$ representing the solvent polarization. This is passed back to the host QC program.
3. The polarization energy $U_{\text{pol}} = \frac{1}{2}(\varphi, \sigma)_\Gamma$ is obtained. This term is the correction to the total energy due to the mutual polarization.
4. The PCM Fock matrix contribution is assembled by contraction of the potential integrals with the solvent polarization:

$$f_{\kappa\lambda}^{\text{PCM}} = \sum_{i=1}^{N_{\text{mesh}}} \sigma(s_i) \int d\mathbf{r} \frac{-\Omega_{\kappa\lambda}(\mathbf{r})}{|\mathbf{r} - s_i|} \qquad (16)$$

5. A new SCF step is performed, a new MEP is obtained and the cycle continues until convergence.

Figure 2 summarizes the algorithm outlined above, highlighting which portions of the program flow are separable between the QC host code and the PCM library.

In the design we have chosen, all operations happening on the PCMSOLVER side only involve functions defined at the cavity boundary, which include, but are not limited to, expectation values of QM quantities, such as the MEP. Even for large systems, such operations are relatively lightweight compared to the integral evaluation and Fock matrix construction. Although not yet implemented in PCMSOLVER, standard techniques in high-performance computing, such as the fast multipole method (FMM) or parallelization, can be employed for very large systems, to reduce the scaling and minimize the computational overhead [124]. The most time-consuming steps for medium to large systems are the calculation of the MEP and assembling the Fock matrix contribution. Their implementation has been left on the host side. There are two clear advantages in using this strategy: on the one hand PCMSOLVER is completely independent of the technology employed on the QM side, keeping the cost of developing the interface minimal; on the other hand it allows the host program to optimize the time-consuming steps without any interference from PCMSOLVER, resulting in optimal performance and minimal computational overhead compared to vacuum calculations.

## 3 | USING THE PCMSOLVER LIBRARY

Avoiding code duplication and encouraging code reuse for common tasks are the main driving forces motivating library writers. Inevitably, libraries evolve over time through trial-and-error. It is expensive and inconvenient to write a software library from a set of written specifications. This is especially true in the computational sciences community, where a consensus on the proper way to acknowledge software output has not yet been reached [6]. Hence one starts from a problem domain and gradually, through refactoring and rewrites, achieves a presumably better API.

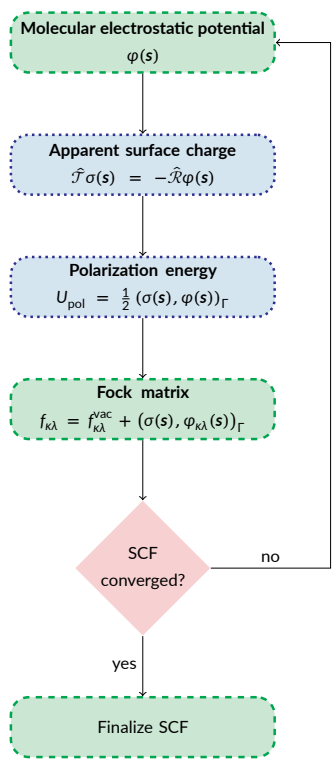

**FIGURE 2** Outline of the SCF algorithm including solvent contributions from the PCM. Blue, dotted outline boxes highlight the operations that are to be implemented in a PCM API. The host QC code will implement the operations and data structures highlighted in the dashed outline green boxes.

PCMSOLVER is written in C++. The object-oriented paradigm provides the necessary flexibility to neatly organize the conceptually different tasks the library has to perform. C++ benefits from a tooling (static and dynamic analysis, linting and style checks) and library ecosystem (chiefly, the standard template library (STL) [76]) that languages such as Fortran have yet to accrue, despite their relatively longer existence. The library also contains Fortran, C and Python components, which we will discuss shortly. CMake is the build system of choice.[1] We adhere to the C++03 ISO standard, which is fully implemented in almost all existing compilers. The GNU, Clang and Intel families of compilers are routinely used with the library for testing and production calculations and are known to work properly. Note that it is still possible to build PCMSOLVER with one of the above mentioned compilers and link it against an executable built with a compiler from another vendor. Dependencies are kept to a minimum and are shipped with the library itself, to minimize the inconvenience for the final users. The C++11 ISO standard introduced new data structures (such as tuples, to model multiple return values from a function), algorithms and tools for functional programming (such as lambdas and argument binding for currying and partial function application) in the core language [88].[2] Our build system is designed to take advantage of these whenever possible and fallback to an alternative implementation in the Boost libraries when an old compiler is used [13]. The library also needs to manipulate vectors and matrices. In the same philosophy of code reuse, we rely on the Eigen C++ template library for linear algebra [55]. Eigen implements containers for vectors and matrices of arbitrary size, both sparse and dense. Operations on these `Eigen::Vector` and `Eigen::Matrix` types are also provided, including a wide array of decompositions and iterative linear solvers. All standard numerical types – integers, single and double precision floating point and their complex counterparts – are supported, with the possibility of extending to custom numerical types. Traditionally, the C and C++ languages have been looked down by the computational science community as offering suboptimal performance in linear algebra operations when compared to Fortran. This can be true with a naïve implementation. Eigen uses *expression templates* and vectorization to overcome this difficulty [142].

---

[1] https://cmake.org/

[2] http://isocpp.github.io/CppCoreGuidelines/CppCoreGuidelines

Writing an interface to PCMSOLVER for your favorite QM code is straightforward. First of all, you will have to download the library. All released versions are available on GitHub, we will refer to the 1.2.1 release (`v1.2.1`) which is the latest version as of this writing. Dependencies and prerequisites are listed on the documentation website and we will assume that all are properly satisfied. Downloading, compiling, testing and installing an optimized version of the library requires few commands:

```
$ curl -L https://github.com/PCMSolver/pcmsolver/archive/v1.2.1.tar.gz | tar -xz
$ cd pcmsolver-1.2.1
# PCMSolver will be built using the Clang C/C++ and GNU Fortran compilers
# with code optimization enabled and installation prefix $HOME/Software
$ ./setup.py --type=release --prefix=$HOME/Software/pcmsolver --cc=clang --cxx=clang++ --fc=gfortran
# Now build with verbose output from compilers and using 2 processes
$ cmake --build build -- VERBOSE=1 -j 2
# Run the full test suite using 2 processes
$ cmake --build build --target test -- -j 2
# We can now install
$ cmake --build build --target install
```

The following installation directory tree will have been generated:

```
$HOME/Software/pcmsolver/
    bin/
        go_pcm.py
        plot_cavity.py
        run_pcm*
    include/
        PCMSolver/
            bi_operators/
            cavity/
            Citation.hpp
            Config.hpp
            Cxx11Workarounds.hpp
            ErrorHandling.hpp
            external/
            green/
            interface/
            LoggerInterface.hpp
            PCMInput.h
            PCMSolverExport.h
            pcmsolver.f90
            pcmsolver.h
            PhysicalConstants.hpp
            solver/
            STLUtils.hpp
            TimerInterface.hpp
            utils/
            VersionInfo.hpp
    lib64/
        libpcm.a
        libpcm.so -> libpcm.so.1*
```

```
            libpcm.so.1*
        python/
            pcmsolver/
    share/
        cmake/
            PCMSolver/
```

The library offers the possibility of saving certain quantities to zipped (.npz) and unzipped (.npy) NumPy binary files for postprocessing and visualization.[3] This requires linking against zlib,[4] which is commonly available on Unix systems. PCMSOLVER includes Fortran components and linking against the Fortran runtime is thus necessary. To summarize, linking your progam to the PCMSOLVER library will require a slight variation on the following commands:

**C/C++ QM host** The program will need to include the header file pcmsolver.h, link against the pcm library (dynamic or static), link against Zlib and the Fortran runtime:

```
# Dynamic linking
$ gcc C_host.c -I. -I$HOME/Software/pcmsolver/include/PCMSolver -o C_host \
    -Wl,-rpath,$HOME/Software/pcmsolver/lib64 $HOME/Software/pcmsolver/lib64/libpcm.so.1
# Static linking
$ gcc C_host.c -I. -I$HOME/Software/pcmsolver/include/PCMSolver -o C_host \
    $HOME/Software/pcmsolver/lib64/libpcm.a -lz -lgfortran -lquadmath -lstdc++ -lm
```

**Fortran QM host** The program will need to compile the pcmsolver.f90 Fortran 90 module source file, link against the pcm library (dynamic or static), link against Zlib and the C++ runtime:

```
# Dynamic linking
$ gfortran $HOME/Software/pcmsolver/include/PCMSolver/pcmsolver.f90 \
        Fortran_host.f90 -o Fortran_host -Wl,-rpath,$HOME/Software/pcmsolver/build/lib64 \
        $HOME/Software/pcmsolver/lib64/libpcm.so.1
# Static linking
$ gfortran $HOME/Software/pcmsolver/include/PCMSolver/pcmsolver.f90 \
        Fortran_host.f90 -o Fortran_host $HOME/Software/pcmsolver/lib64/libpcm.a -lstdc++ -lz
```

These build requirements for the QM host program can be managed within a Makefile. For host programs using CMake, a configuration file is also provided such that a find_package(PCMSolver) directive will search for the library and import all that is necessary to link.

Once the linking issues are sorted out, the QM code will need a function[5] to compute the MEP on a grid of points. The signature for such a function might look as follows:

```
! Calculate electrostatic potential
! φ_i ≡ φ(s_i) = Σ_{A=1}^{N_nuclei} Z_A/|R_A−s_i| + Σ_{κλ} D_{κλ} ∫ dr (−Ω_{κλ}(r))/|r−s_i|, ∀i = 1, N_mesh
pure subroutine get_mep(nr_nuclei, nuclear_charges, nuclear_coordinates, density_matrix, nr_mesh, grid,
    ↪ mep)
  implicit none
  use iso_fortran_env, only: int32, real64
  integer(int32), intent(in) :: nr_nuclei
  real(real64),   intent(in) :: nuclear_charges(nr_nuclei)
```

---

[3]https://github.com/rogersce/cnpy, https://docs.scipy.org/doc/numpy/neps/npy-format.html

[4]https://zlib.net/

[5]We will use the term "function" throughout, even though Fortran has a distinction between a subroutine (in C parlance, a function that *does not* return, *i.e.* void a_subroutine) and a function (in C parlance, a function that *does* return, *i.e.* double a_function).

```fortran
    real(real64),    intent(in) :: nuclear_coordinates(3, nr_nuclei)
    real(real64),    intent(in) :: density_matrix(*)
    integer(int32),  intent(in) :: nr_mesh
    real(real64),    intent(in) :: grid(3, nr_mesh)
    real(real64), intent(inout) :: mep(nr_mesh)
  end subroutine
```

A function to compute the PCM contribution to the Fock matrix (or to the $\sigma$-vector in response theory) is also needed. This is a modified one-electron nuclear attraction potential and a possible signature is as follows:

```fortran
! Calculate contraction of apparent surface charge with charge-attraction integrals
! f_{\kappa\lambda}^{PCM} = (\sigma(s), \varphi_{\kappa\lambda}(s))_\Gamma \equiv \sum_{i=1}^{N_{mesh}} \sigma(s_i) \int dr \frac{-\Omega_{\kappa\lambda}(r)}{|r-s_i|}
pure subroutine get_pcm_fock(nr_mesh, asc, fock_matrix)
  implicit none
  use iso_fortran_env, only: int32, real64
  integer(int32),  intent(in) :: nr_mesh
  real(real64),    intent(in) :: asc(nr_mesh)
  real(real64), intent(inout) :: fock_matrix(*)
end subroutine
```

These functions *are not provided by* PCMSOLVER. Indeed, the library has been designed based on the realization that the PCM layer is *completely* independent of the AO or MO spaces defined in the quantum chemical layer. As discussed in section 2.3 and schematically shown in figure 3, there is no need for the PCM library to handle integrals, density and Fock matrices. This architecture avoids handling large data structures, such as the density and Fock matrices, and code duplication at the integral computation level. In addition, it makes PCMSOLVER *fully agnostic* of the QM host program: no assumptions are made on the storage format for matrices or the way AO basis integrals are computed. This is the main strength of PCMSOLVER and has led to its inclusion into many different QM host programs with negligible computational overhead.

Initialization of the library happens with a call to the `pcmsolver_new` function. This function returns a *context*, which will be the handle to any PCM-related operations in the rest of the calculation.

```fortran
interface pcmsolver_new
  function pcmsolver_new(input_reading, nr_nuclei, charges, coordinates, symmetry_info, host_input,
  ↪ writer) result(context) bind(C)
    import
    integer(c_int), intent(in), value :: input_reading
    integer(c_int), intent(in), value :: nr_nuclei
    real(c_double), intent(in)        :: charges(*)
    real(c_double), intent(in)        :: coordinates(*)
    integer(c_int), intent(in)        :: symmetry_info(*)
    type(PCMInput), intent(in)        :: host_input
    type(c_funptr), intent(in), value :: writer
    type(c_ptr) :: context
  end function
end interface
```

The `pcmsolver_new` function requires the number of atomic centers `nr_nuclei`, their charges and coordinates `coordinates`, the symmetry generators `symmetry_info` (Abelian groups only are supported) and a function pointer `writer` to output facilities within the host program. The additional parameters to the function are needed to handle PCM-specific input. Currently, the module can either read its own input file from disk or from the `host_input` data structure as filled by the host program. This design choice was made to allow for a fast initial implementation of PCM within a host program, one that would not require extensive reorganization of the host program's own input parsing functions. The trade-off is that the user now has to make sure that the PCMSOLVER input is parsed and the resulting intermediate, machine-readable file is available at run-time in the appropriate directory. We provide the `go_pcm.py` Python script for this purpose, which parses and validates the input file by means of the GetKw library [77]. A PCMSOLVER input file is organized into *keywords* and *sections*, which are collections of keywords. Each section roughly maps to a computational task in the library: how to build the cavity, what Green's function to use and how to set up the solver. The following sample input asks for a conductor-like polarizable continuum model (CPCM) calculation with methanol as a solvent:

```
units = angstrom
cavity
{
  type = gepol
  area = 0.6
  mode = atoms
  atoms = [1, 4]
  radii = [1.2, 1.8]
}

medium
{
  solvent = methanol
  solvertype = cpcm
  correction = 0.5
  diagonalscaling = 1.0694
}
```

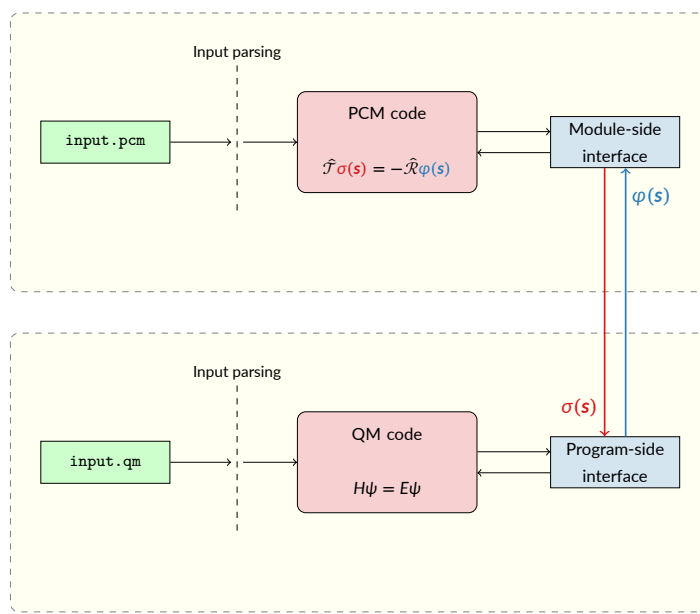

**FIGURE 3** High-level view of the relationship between a host quantum chemistry program and the PCMSOLVER library. The initialization phase, represented by the input parsing portions, will generate the molecular cavity and the PCM matrix for the chosen environment. During the iterative solution of the Schrödinger equation, *by any method*, the MEP, $\varphi(s)$, and ASC, $\sigma(s)$, are the only data to be passed back and forth between library and host code. This affords a significant streamlining of the interfaces to be written.

The average area of the generated finite elements will be 0.6 Å$^2$ (or less), spheres will be put on all atoms, with the radii for the first and fourth in the list passed from the host program will have a custom-set radius. The CPCM solver will be set up with a dielectric scaling of $f(\varepsilon) = \frac{\varepsilon-1}{\varepsilon+0.5}$, the diagonal elements of the boundary integral operators $\hat{S}$ and $\hat{D}$ will be scaled by the given factor of 1.0694. At initialization, the library will generate the cavity, set up the Green's functions, compute the boundary integrals operators and assemble the solver. All further interactions between the host program and PCMSOLVER happen through the `context` pointer returned by the `pcmsovler_new` function, that is, the first argument in all API function is the PCM context. This allows for *more than one* PCM object existing at once during a calculation, each with its separate set up, an idea akin to the *execution plans* in the FFTW3 library [51].

The next step is the calculation of the MEP at the cavity grid points. The QM host fetches the size of the grid with the `pcmsolver_get_cavity_size` function:

```fortran
interface pcmsolver_get_cavity_size
  function pcmsolver_get_cavity_size(context) result(nr_points) bind(C)
    import
    type(c_ptr), value :: context
    integer(c_int)  :: nr_points
  end function
end interface
```

allocates memory accordingly and fetches the grid by calling `pcmsolver_get_centers`:

```fortran
interface pcmsolver_get_centers
  subroutine pcmsolver_get_centers(context, centers) bind(C)
    import
    type(c_ptr), value :: context
    real(c_double), intent(inout) :: centers(*)
  end subroutine
end interface
```

The QM host code can decide whether to save the PCM grid in memory (globally or in a data structure localized to the SCF portion of the code), on disk or repeatedly calling the `pcmsolver_get_centers` function when needed. After calling the relevant integral evaluation functions, the MEP will be available as a vector of size equal to that of the cavity mesh. When uniquely labeled, say `TotMEP` for the MEP, we refer to such quantities as *surface functions*. The PCM context holds a collection of (`label`, `data`) pairs of such functions, what is called an *associative array*, *dictionary* or *map*. The host program can set and get surface functions with the appropriate functions. The functionality has been programmed to avoid unnecessary copies of the data and to allow for arbitrary labels for the functions. During an SCF iteration we add, or modify the contents of, the MEP surface function by calling `pcmsolver_set_surface_function` with our label of choice:

```fortran
interface pcmsolver_set_surface_function
  subroutine pcmsolver_set_surface_function(context, f_size, values, name) bind(C)
    import
    type(c_ptr), value :: context
    integer(c_int), value, intent(in) :: f_size
    real(c_double), intent(in) :: values(*)
    character(kind=c_char, len=1), intent(in) :: name(*)
  end subroutine
end interface
```

Everything is now in place to compute the ASC. Much as the MEP, the ASC is also a surface function. For its computation the `pcmsolver_compute_asc` function is provided:

```fortran
interface pcmsolver_compute_asc
  subroutine pcmsolver_compute_asc(context, mep_name, asc_name, irrep) bind(C)
    import
    type(c_ptr), value :: context
    character(kind=c_char, len=1), intent(in) :: mep_name(*), asc_name(*)
    integer(c_int), value, intent(in) :: irrep
  end subroutine
end interface
```

accepting *two* surface function labels. PCMSOLVER will compute the ASC using the requested solver and create, or update, the corresponding entry in the surface function dictionary. The host program can then retrieve the ASC invoking `pcmsolver_get_surface_function`:

```fortran
interface pcmsolver_get_surface_function
  subroutine pcmsolver_get_surface_function(context, f_size, values, name) bind(C)
    import
    type(c_ptr), value :: context
    integer(c_int), value, intent(in) :: f_size
    real(c_double), intent(inout) :: values(*)
    character(kind=c_char, len=1), intent(in) :: name(*)
  end subroutine
end interface
```

in a fashion that is symmetric to the `pcmsolver_set_surface_function`. We remark once again that data transfer between PCMSOLVER and the QM host program is limited to the communication of $\{\varphi(s_i)\}_{i=1}^{N_{\mathrm{mesh}}}$ and $\{\sigma(s_i)\}_{i=1}^{N_{\mathrm{mesh}}}$ and is implemented *without* storing any quantity to disk, avoiding any overhead I/O operations might incur. The correction, $U_{\mathrm{pol}}$, to the total energy due to the polarization of the continuum can be calculated as the dot product of the MEP and ASC arrays. PCMSOLVER also provides a function, `pcmsolver_compute_polarization_energy`, with a signature similar to that of `pcmsolver_compute_asc`

```fortran
! Compute U_pol = 1/2 (σ, φ)_Γ ≡ 1/2 Σ_{i=1}^{N_mesh} σ(s_i)φ(s_i)
interface pcmsolver_compute_polarization_energy
  function pcmsolver_compute_polarization_energy(context, mep_name, asc_name) result(energy) bind(C)
    import
    type(c_ptr), value :: context
    character(kind=c_char, len=1), intent(in) :: mep_name(*), asc_name(*)
    real(c_double) :: energy
  end function
end interface
```

The PCM contribution to the Fock matrix can now be computed by calling the appropriate function in the QM host program. Listing 1 summarizes the steps necessary to get SCF up and running including the PCM solvent contributions.

**LISTING 1** Pseudocode summary of the calls to achieve SCF iterations including PCM contributions in a Fortran code. Full working examples are available in the PCMSOLVER online repository for a C host and a Fortran host.

```fortran
program pcm_fortran_host
  use, intrinsic :: iso_c_binding
  use, intrinsic :: iso_fortran_env, only: output_unit, error_unit
  use pcmsolver
  implicit none
  integer(c_int) :: nr_nuclei ! Number of atomic centers
  real(c_double), allocatable :: charges(:) ! Atomic charges
  real(c_double), allocatable :: coordinates(:) ! Coordinates of the atomic centers, a (3, nr_nuclei) array in column-major order
  type(c_ptr) :: pcm_context ! Handle to the PCMSolver library
  integer(c_int) :: symmetry_info(4) ! Point group symmetry generators
  type(PCMInput) :: host_input ! Input reading data structure
  character(kind=c_char, len=*), parameter :: mep_lbl = 'TotMEP' ! Molecular electrostatic (MEP) potential surface function label
  character(kind=c_char, len=*), parameter :: asc_lbl = 'TotASC' ! Apparent surface charge (ASC) surface function label
  integer(c_int) :: grid_size ! The PCM cavity mesh grid size
  real(c_double), allocatable :: grid(:) ! The PCM cavity mesh coordinates, a (3, grid_size) array in column-major order
  real(c_double), allocatable :: mep(:), asc(:) ! The MEP and ASC arrays
  real(c_double) :: Upol ! The polarization energy
  ! Input parsing for QM code and initialize QM code internals
  nr_nuclei = get_nr_nuclei()
  allocate(charges(nr_nuclei))
  allocate(coordinates(3*nr_nuclei))
  call get_molecule(nr_nuclei, charges, coordinates)
  ! Initialize PCMSolver. It is assumed that parsing of the PCM input has already happened
  if (.not. pcmsolver_is_compatible_library()) then
     write(error_unit, *) 'PCMSolver library not compatible!'
     stop
  end if
  ! symmetry_info, host_input and host_writer are here assumed to have been initialized
  pcm_context = pcmsolver_new(PCMSOLVER_READER_HOST, nr_nuclei, charges, coordinates, symmetry_info, host_input, c_funloc(host_writer))
  call pcmsolver_print(pcm_context) ! Print PCMSolver set up information
  grid_size = pcmsolver_get_cavity_size(pcm_context) ! Get size of the PCM cavity mesh
  allocate(grid(3*grid_size)) ! Allocate space for the PCM cavity mesh coordinates
  grid = 0.0_c_double
  call pcmsolver_get_centers(pcm_context, grid) ! Get the PCM cavity mesh
  !!! SCF iterations !!!
  ! Calculate and set TotMEP surface function
  allocate(mep(grid_size))
  mep = 0.0_c_double
  call get_mep(nr_nuclei, charges, coordinates, density_matrix, grid_size, grid, mep)
  call pcmsolver_set_surface_function(pcm_context, grid_size, mep, pcmsolver_fstring_to_carray(mep_lbl))
  ! Compute the ASC surface function for the totally symmetric irrep
  call pcmsolver_compute_asc(pcm_context, pcmsolver_fstring_to_carray(mep_lbl), pcmsolver_fstring_to_carray(asc_lbl), irrep = 0_c_int)
  ! Grab the ASC surface function into the appropriate array
  allocate(asc(grid_size))
  asc = 0.0_c_double
  call pcmsolver_get_surface_function(pcm_context, grid_size, asc, pcmsolver_fstring_to_carray(asc_lbl))
  energy = pcmsolver_compute_polarization_energy(pcm_context, mep_lbl, asc_lbl)
  write(output_unit, '(A, F20.12)') 'Polarization energy = ', energy
  ! Calculate contraction of apparent surface charge with charge-attraction integrals
  call get_pcm_fock(grid_size, asc, fock_matrix)
  !!! End of SCF iterations !!!
  call pcmsolver_save_surface_functions(pcm_context) ! Save converged surface functions to NumPy arrays
  ! Clean up MEP and ASC arrays
  deallocate(mep)
  deallocate(asc)
  ! Finalize PCMSolver library
  call pcmsolver_delete(pcm_context)
  ! Clean up PCM cavity mesh coordinates array
  deallocate(grid)
  deallocate(charges)
  deallocate(coordinates)
  close(output_unit)
end program
```

# 4 | DEVELOPING THE PCMSOLVER LIBRARY

Grasping the inner workings of an unfamiliar piece of software is always difficult and the aim of this section is to minimize this effort for potential new contributors to the PCMSOLVER library. It will not be possible to give an explanation in full detail of all of our design choices and motivations, but this will constitute a good primer. Whereas section 3 provided a top-down description of the library, this section will offer the complementary bottom-up view. PCMSOLVER is written in a combination of well-established compiled languages C++, C and Fortran with additional tooling provided by Python scripts and modules. Cloning the PCMSOLVER Git repository will generate the following directory layout:

```
pcmsolver/
    api/                    # API functions
    cmake/                  # CMake modules
    doc/                    # reStructuredText documentation sources
    examples/               # Sample inputs
    external/               # Prepackaged external dependencies
    include/                # Library internal header files
    src/                    # Library internal source files
        bin/                # Standalone executable for testing
        bi_operators/       # Computation of boundary integral operators
        cavity/             # Cavity definition and meshing
        green/              # Green's functions definitions
        interface/          # API-internals
        pedra/              # GEPOL cavity generator
        solver/             # Integral equation set up and solution
        utils/              # General purpose utilities
    tests/                  # Unit tests and API tests
    tools/                  # Python tools
```

Figure 4 shows basic statistics about the source code repository.

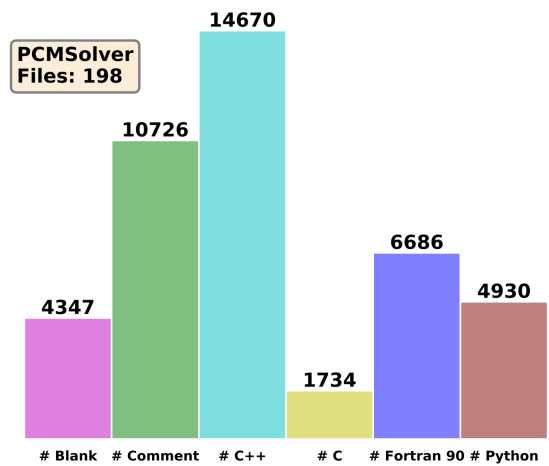

**FIGURE 4** Number of source files and lines of code (LOC) statistics for PCMSOLVER. The LOC count is broken down by language. The comments include Doxygen markup for the autogenerated class and function documentation.

Solving the boundary integral equation (BIE) (3) by means of the BEM requires a number of ingredients: a boundary mesh generator, computational kernels for the Green's functions, backends for the computation of the discretized boundary integral operators and finally a linear system solver.[6] The geography of these ingredients in PCMSOLVER is as follows:

**Mesh generator: folders `cavity` and `pedra`** Different BEM methods might pose different constraints for the generator. For example, triangular *vs.* quadrilateral or planar *vs.* spherical patches. All these points have been discussed at length in the BEM and PCM literatures [43, 104] and we will briefly review the available mesh generator in PCMSOLVER.

**Green's functions: folder `green`** Depending on the nature of the BIE, up to second order derivatives of the Green's function might be needed to set up the boundary integral operators. The IEF-PCM equation (3) only requires the conormal derivatives, however the breadth of Green's functions currently implemented in PCMSOLVER (see Table 1) poses a challenge for the implementation of this component. We shall show that automatic differentiation (AD) [21] in combination with static (template-based) *and* dynamic (class-based) polymorphism [116, 142] provides a robust, clean and extensible framework for implementing Green's functions and their derivatives.

**Computation of the BI operators on the mesh: folder `bi_operators`** As discussed in section 2.1, the integrals needed are multidimensional and on possibly arbitrary domain shapes. On top of these difficulties, the operators are also singular. Techniques and algorithms have been developed and the interested reader can refer to the monograph by Sauter and Schwab [123]. The library implements a straightforward collocation scheme which we will not discuss in further detail.

**PCM equation solver: folder `solver`** The solver can be direct or iterative, the latter even in a matrix-free flavor. PCMSOLVER uses the stock implementation in Eigen of standard algorithms [55, 54]. For CPCM the $\hat{\mathcal{S}}$ matrix is stored and a Cholesky decomposition is used:

```
Eigen::VectorXd ASC = -S_.llt().solve(MEP);
```

For IEF-PCM the $\hat{\mathcal{T}}$ and $\hat{\mathcal{R}}$ matrices are stored and a partially pivoted **LU** decomposition is used. By default, we compute polarization weights, requiring the solution of *two* linear systems of equations per call [30]:

```
// ASC: σ = -T̂⁻¹R̂φ
Eigen::VectorXd ASC = - T_.partialPivLu().solve(R_ * MEP);
// Adjoint ASC: σ* = -R̂†(T̂†)⁻¹φ
// First compute χ = (T̂†)⁻¹φ, then compute σ* = -R̂†χ
Eigen::VectorXd adj_ASC = T_.adjoint().partialPivLu().solve(MEP);
adj_ASC = -R_.adjoint() * adj_ASC.eval();
// Get polarization weights: ω = ½(σ + σ*)
ASC = 0.5 * (adj_ASC + ASC.eval());
```

The user can turn off the computation of the polarization weights by setting `hermitivitize=false` in the input, though this is not recommended.[7]

Finally, the `interface` folder contains the `Meddle` class which orchestrates the initialization/finalization of the library and the computation of the ASC. This is the backend for the API functions defined in the `pcmsolver.h` header file and exported to Fortran in the `pcmsolver.f90` module source file. These latter files are contained in the `api` folder.

The internal structure of the library is shown in figure 5 in relation with the API functions discussed in section 3. The green layer at the bottom of the figure shows the dependencies of PCMSOLVER:

---

[6] Many of the same ingredients are shared with Finite Element Method (FEM) codes.

[7] In our experience the use of polarization weights helps SCF convergence and is essential for a stable iterative solution of the linear equations arising in response theory.

- Eigen: a C++ template library for linear algebra [55].
- libtaylor: a C++ template library [42] for AD [21].
- libgetkw: a library for input parsing [77].
- Boost: a general purpose C++ library [13]. In PCMSOLVER it provides the ODE integrator [15] and the C++11 compatibility layer for older compilers.

These dependencies are included with the source code repository, but are only used in the building process if proper versions are not found preinstalled on the system. Users need not worry about satisfying dependencies beforehand. This makes PCMSOLVER a self-contained, but somewhat heavy library. The yellow layer contains the heavy-lifting

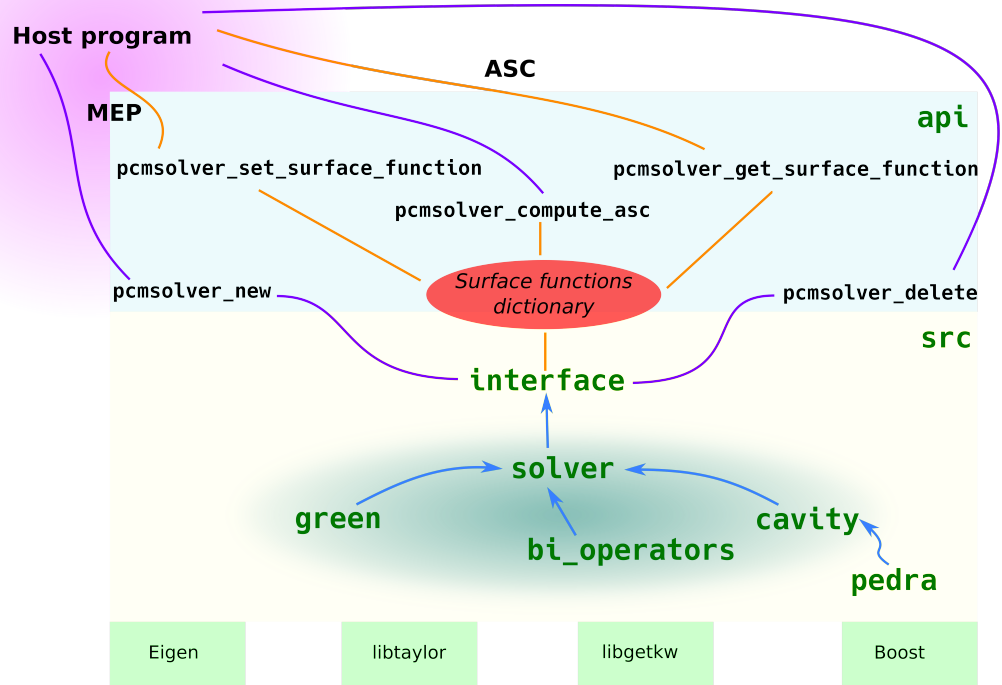

**FIGURE 5** Internal structure of the PCMSOLVER library in relation to the API and the host program. The green boxes at the bottom show the external dependencies. The internal implementation of the API is contained in the `src` folder and is shown in the yellow layer. The blue arrows exemplify the composition relations between the data structures defined in each folder. The upper, blue layer is the exposed API of the PCMSOLVER library. The initialization (`pcmsolver_new`), finalization (`pcmsolver_delete`) and surface function manipulation functions (`pcmsolver_get_surface_function`, `pcmsolver_set_surface_function`, `pcmsolver_compute_asc`) and their relation with the host program and the API internals, defined in the `interface` folder, are shown. Orange lines show the flow of data between these components, whereas the purple lines show the control flow.

portions of the library, which maps to the contents of the `src` folder.

**Cavity generation**

Building the molecular cavity is the starting point, a task accomplished by sources in the `cavity` and `pedra` folders. In continuum solvation models (CSMs) it is almost always the union of a set of spheres centered on the atoms.[8] The atomic radii used vary wildly among different implementations. Possible choices implemented in PCMSOLVER are: van der Waals radii as tabulated by Bondi [22] (and later extended by Mantina et al. [90]), the UFF radii [115] or the set of Allinger et al. [18]. Once sphere centers and radii are settled upon, one has the *van der Waals surface*, $S_{\mathrm{vdW}}$. This might be too tight, what is usually done is a rescaling of the radii by a factor $\alpha = 1.2$. We also want the definition of molecular surface to capture the fact that solvent molecules cannot penetrate within the molecule of interest. The *solvent-accessible surface* – $S_{\mathrm{SAS}}$ – is defined as the locus of points described by the *center* of a spherical probe, modeling a solvent molecule, rolling on $S_{\mathrm{vdW}}$. The *solvent-excluded surface* – $S_{\mathrm{SES}}$ – instead is the locus of points described by the *contact point* of a spherical probe rolling on the $S_{\mathrm{vdW}}$. Whereas $S_{\mathrm{vdW}}$ and $S_{\mathrm{SAS}}$ only consist of convex spherical patches, $S_{\mathrm{SES}}$ consists of convex and concave spherical and toroidal patches [28, 57, 113, 114]. To ensure continuity of energy gradients, this union of spheres can be smoothed [146, 107, 134, 125, 80, 81, 138]. The implementation of efficient meshing algorithms for the boundary surfaces defined above is still a very active area of research. PCMSOLVER offers the venerated GEPOL (*GE*nerating *POL*yhedra) algorithm, first devised in the 80s [101] and gradually improved [130, 103, 131, 102, 106, 108]. GEPOL approximates $S_{\mathrm{SES}}$ by adding spheres not centered on atoms to fill up the portions of space where the solvent cannot penetrate, the mesh generation starts from a set of equilateral triangles defined by the vertices of a regular polyhedron inscribed in the spheres. The spherical triangles are then cut at the spheres intersection. An iterative refinement, by successively cutting into smaller triangles, is performed until the average area of the finite elements reaches a predefined user threshold. The Fortran implementation of GEPOL (folder `pedra`) is wrapped into a C++ container class `GePolCavity`. The container class holds all the data produced by the meshing algorithm: collocation points (centroids of the finite elements), weights (areas of the finite elements), outward pointing normal vectors, curvature, arcs and vertices. These data are saved to a compressed NumPy array (`.npz` format) for postprocessing in Python, see figure 6. The GEPOL algorithm has some well-known shortcomings [104] and an implementation of the TsLess algorithm of Pomelli [107] is currently underway.[9]

**Green's functions**

Green's functions are the next basic building block in our hierarchy. Given Eq. (6) for the general form a Green's function, we have implemented the following type:

```
class IGreensFunction {
public:
  /*! Returns value of the kernel of the Ŝ integral operator for the pair
   * of points p₁, p₂: G(p₁,p₂)
   */
  virtual double kernelS(const Eigen::Vector3d & p1,
              const Eigen::Vector3d & p2) const = 0;
  /*! Returns value of the kernel of the D̂ integral operator for the
   * pair of points p₁, p₂: [ε∇_{p₂}G(p₁,p₂)]·n_{p₂}
   */
  virtual double kernelD(const Eigen::Vector3d & direction,
              const Eigen::Vector3d & p1,
```

---

[8] Notable exceptions are the DefPol [105, 109] and the isodensity PCM algorithms [46, 44, 126, 19, 47, 45].
[9] Work-In-Progress pull request on GitHub: https://github.com/PCMSolver/pcmsolver/pull/140

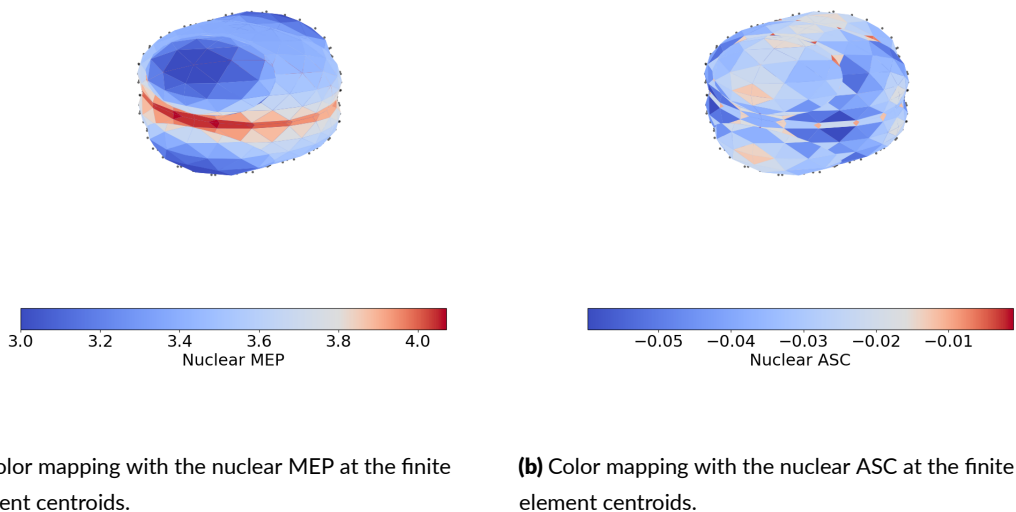

(a) Color mapping with the nuclear MEP at the finite element centroids.

(b) Color mapping with the nuclear ASC at the finite element centroids.

**FIGURE 6** The GEPOL cavity for the ethene molecule in $C_1$ symmetry. The finite element centroids are represented by dots. The figure was obtained from the `cavity.npz` and MEP and ASC NumPy array file produced by PCMSOLVER and the `plot_cavity.py` script. Color bars in atomic units. Water ($\varepsilon$= 78.39) was selected as solvent.

```
                    const Eigen::Vector3d & p2) const = 0;

  /*! Calculates an element on the diagonal of the matrix representation of the
   * Ŝ operator using an approximate collocation formula.
   */
  virtual double singleLayer(const Element & e, double factor) const = 0;
  /*! Calculates an element of the diagonal of the matrix representation of the D̂
   * operator using an approximate collocation formula.
   */
  virtual double doubleLayer(const Element & e, double factor) const =0;
};
```

The *pure virtual methods* (`virtual ... = 0;`) mean that this type is *abstract*, providing the definition of an interface. It carries no information whatsoever regarding *how* to compute the value of a Green's function, it only prescribes what kind of operations a *concrete* Green's function type has to explicitly implement to be valid [52, 88]. These are:

- The `kernelS` function for the calculation of its value, given a pair of points in space.
- The `kernelD` function for the calculation of its directional derivative, given a pair of points in space and a direction.
- The `singleLayer` function, for the calculation of the *diagonal* elements of the $\hat{S}$ operator given a finite element.
- The `doubleLayer` function, for the calculation of the *diagonal* elements of the $\hat{\mathcal{D}}$ operator given a finite element.

Concrete types for Green's functions, for example a type for the uniform dielectric or the spherical sharp, will have to conform to this interface so that we will be able to produce a valid boundary integral operator with the same set of commands. For example, the $\hat{\mathcal{T}}$ operator for the anisotropic IEF-PCM equation is assembled from the cavity boundary,

$G_i$, $G_e$ and a boundary integral operator engine as follows:

```cpp
Eigen::MatrixXd anisotropicTEpsilon(const ICavity & cav,
                                    const IGreensFunction & gf_i,
                                    const IGreensFunction & gf_o,
                                    const IBoundaryIntegralOperator & op) {
  Eigen::MatrixXd SI = op.computeS(cav, gf_i);
  Eigen::MatrixXd DI = op.computeD(cav, gf_i);
  Eigen::MatrixXd SE = op.computeS(cav, gf_o);
  Eigen::MatrixXd DE = op.computeD(cav, gf_o);
  Eigen::MatrixXd a = cav.elementArea().asDiagonal();
  Eigen::MatrixXd Id = Eigen::MatrixXd::Identity(cav.size(), cav.size());
  Eigen::MatrixXd T = ((2 * M_PI * Id - DE * a) * SI +
                      SE * (2 * M_PI * Id + a * DI.adjoint().eval()));
  return T;
}
```

PCMSOLVER uses *forward-mode* AD implemented through *operator overloading* to obtain the necessary derivatives [21]. In forward-mode AD, basic data types are augmented by incorporating an infinitesimal component $\epsilon$: $x := x + x'\epsilon$ where the coefficient $x'$ is the value of the derivative at the given point. Arithmetic operators and elementary functions are then redefined to accept these augmented types. Composition of elementary functions maps to the application of the chain rule for derivatives. Evaluating an arbitrarily complex function composed from these primitives yields the value of the function itself and of its derivatives *to any order* with the same *numerical accuracy*. AD sidesteps the problems inherent to numerical differentiation. The additional programming effort is also reduced with respect to an analytic implementation of the derivatives. PCMSOLVER uses libtaylor [42] which implements forward-mode AD for arbitrary multivariate functions and derivative orders. This is achieved by means of a type, `taylor<T, V, D>`, storing the Taylor expansion coefficients of a `V`-variate function as a `V`-variate polynomial of numeric type `T` and degree `D`. All the elementary functions, arithmetic and ordering operators available in C++ are redefined for this type by libtaylor. The use of template programming guarantees the open-endedness in terms of the underlying type, degree and number of variables.[10] To combine the benefits of forward-mode AD with our code, the concrete Green's functions types have to be *parametrized over* the `taylor<T, V, D>` type of choice, for directional derivatives of Green's function the type `taylor<double, 1, 1>` is sufficient. Listing 2 shows a skeleton implementation of the Green's function for the uniform dielectric.[11] The concrete class `UniformDielectric` is parametrized over a `taylor` type (template-based *static* polymorphism) *and* inherits from the abstract base class (class-based *dynamic* polymorpshim) to implement the operations outlined by the base class using AD [52, 17, 135]. The function call operator, `operator()`, is where the Green's function is defined, thanks to AD the return value will also contain its directional derivative. We can thus outline an implementation checklist for Green's functions, valid also for more complicated environments:

1. Define the input parameters for the Green's function (*e.g.* permittivity $\varepsilon$) and write a *constructor* function to initialize the data from a passed value. The construction phase can be arbitrarily complex. For example, the diffuse interface in spherical symmetry requires the solution of sets of radial ODEs.
2. Provide an implementation for the function call operator `operator()`, returning the value of the Green's function. `operator()` can be arbitrarily complex: the sharp interface in spherical symmetry has to implement the separation of the Coulomb and image components and calculate the latter as a truncated sum over Legendre polynomials.

---

[10] For further details consult the source code available on GitHub: https://github.com/uekstrom/libtaylor

[11] The actual implementation in PCMSOLVER is slightly more involved. The permittivity of the environment is modeled as a profile function (sharp, diffuse, anisotropic and so forth) which becomes a template parameter of the concrete class implementing the Green's function of choice.

**LISTING 2** Skeleton of the implementation of the uniform dielectric Green's function.

```cpp
template <typename DerivativeTraits = taylor<double, 1, 1> >
class UniformDielectric : public IGreensFunction {
public:
  // Constructor: initializes a uniform dielectric Green's function given a permittivity
  UniformDielectric(double eps) : epsilon_(eps) {}
  // Implements the pure virtual kernelS function
  virtual double kernelS(const Eigen::Vector3d & p1, const Eigen::Vector3d & p2)
  const {
    DerivativeTraits sp[3], pp[3];
    sp[0] = p1(0);
    sp[1] = p1(1);
    sp[2] = p1(2);
    pp[0] = p2(0);
    pp[1] = p2(1);
    pp[2] = p2(2);
    return this->operator()(sp, pp)[0];
  }
  // Implements the pure virtual kernelD function
  virtual double kernelD(const Eigen::Vector3d & direction,
                         const Eigen::Vector3d & p1,
                         const Eigen::Vector3d & p2) const {
    DerivativeTraits t1[3], t2[3];
    t1[0] = p1(0);
    t1[1] = p1(1);
    t1[2] = p1(2);
    t2[0] = p2(0);
    t2[1] = p2(1);
    t2[2] = p2(2);
    t2[0][1] = normal_p2(0);
    t2[1][1] = normal_p2(1);
    t2[2][1] = normal_p2(2);
    return this->operator()(t1, t2)[1];
  }
  // Implements the pure virtual singleLayer function
  virtual double singleLayer(const Element & e, double factor) const {
    return (factor * std::sqrt(4 * M_PI / area));
  }
  // Implements the pure virtual doubleLayer function
  virtual double doubleLayer(const Element & e, double factor) const {
    return (-factor * std::sqrt(M_PI / area) * 1.0 / radius);
  }
private:
  // Permittivity ε of the uniform dielectric
  double epsilon_;
  // Function call operator computing the value of the function and its
  // derivatives, as prescribed by the DerivativeTraits type.
  // The DerivativeTraits type defaults to the directional derivative, that is
  // of type taylor<double, 1, 1>
  DerivativeTraits operator()(DerivativeTraits * sp,
                              DerivativeTraits * pp) const {
    return 1 / (this->epsilon_ * distance(sp, pp));
  }
};
```

3. Implement the `kernelS` and `kernelD` methods, in terms of `operator()`.[12]
4. Implement the `singleLayer` and `doubleLayer` methods.

**The interface**

The API of PCMSOLVER is implemented in ISO C99. The functions we described in section 3 call a corresponding method in the `Meddle` class, defined in the `interface` folder. The API is *context-aware* [121]. Initialization of the library *via* the `pcmsolver_new` function creates all objects relevant to the calculation and return a handle to the library in the form of a context object.[13] The context object is an opaque C `struct`: a pointer to some other object, in our case an instance of the `Meddle` class owning the current calculation set up.

When using dynamic polymorphism, instances of concrete classes are used through pointers to their corresponding abstract base classes (Liskov substitution principle [52]). For example, the following declares a vacuum and a uniform dielectric (water) Green's functions, with derivatives calculated using AD.

```
IGreensFunction * gf_i = new Vacuum<>();
IGreensFunction * gf_o = new UniformDielectric<>(78.39);
```

PCMSOLVER offers quite a number of knobs to tune the set up of a calculation. Naïve approaches to the initialization might lead to poor design choices, like a nested, factorial branching logic or the use of type casting.[14] We have adopted the Factory method pattern, a standard solution that avoids both pitfalls [52, 17]:[15]

```
IGreensFunction * gf_i = green::bootstrapFactory().create(
    input_.insideGreenParams().greensFunctionType, input_.insideGreenParams());
IGreensFunction * gf_o = green::bootstrapFactory().create(
    input_.outsideStaticGreenParams().greensFunctionType,
    input_.outsideStaticGreenParams());
```

The usage of a context-aware API hides many implementation details of the PCM from the host QM code. For example, this is the body of the `pcmsolver_compute_asc` function and its counterpart in the `Meddle` object:

```
void pcmsolver_compute_asc(pcmsolver_context_t * context,
                           const char * mep_name,
                           const char * asc_name,
                           int irrep) {
  reinterpret_cast<pcm::Meddle *>(context)->computeASC(std::string(mep_name), std::string(asc_name),
    ↪ irrep);
}
void pcm::Meddle::computeASC(const std::string & mep_name,
                             const std::string & asc_name,
```

---

[12] We note that deferring the implementation of the `kernelS` and `kernelD` methods to the concrete classes leads to a lot of boilerplate, error-prone code. In our current implementation, this is avoided by providing these methods in an intermediate template class `template <typename DerivativeTraits, ProfilePolicy> class GreensFunction`. This approach is admittedly more involved, but reduces code duplication and allows us to neatly include the corner case where numerical differentiation of the Green's function is desired or necessary.

[13] https://github.com/bast/context-api-example

[14] The use of `dynamic_cast` allows casting up and down an inheritance hierarchy, thus deferring the creation of the concrete type until it's properly localized. C++ however does not have introspection and using the `dynamic_cast` construct introduces a run-time performance penalty. Apart from this, it also completely bypasses the type system, thus nullifying the benefits of inheritance hierarchies and strong typing.

[15] We have a template implementation that follows the one presented by Alexandrescu [17]. The factory stores an associative container (`std::map`) of object tags and callback creation functions. When calling the `create` method, the container is traversed to find the tag and the corresponding callback is invoked. The arguments and return type of the callback are deduced by the compiler. Traversal of a `std::map` to obtain the correct callback function can be *more efficient* than branching, even when only few conditional branches would be needed [17, 76].

```
                        int irrep) const {
  // Get the proper iterators
  SurfaceFunctionMapConstIter iter_pot = functions_.find(mep_name);
  Eigen::VectorXd asc = K_0_->computeCharge(iter_pot->second, irrep);
  // Renormalize for the number of irreps in the group
  asc /= double(cavity_->pointGroup().nrIrrep());
  // Insert it into the map
  if (functions_.count(asc_name) == 1) { // Key in map already
    functions_[asc_name] = asc;
  } else { // Create key-value pair
    functions_.insert(std::make_pair(asc_name, asc));
  }
}
```

## 5 | CONTRIBUTING TO PCMSOLVER

PCMSOLVER is released under the terms of the GNU Lesser General Public Licence, version 3, a standard open-source license.[16] The LGPL is a weak-copyleft license [122, 133]. It is well-suited for the open-source distribution of libraries, since it strikes a balance between openness and protection of the ideas implemented in the distributed code. The LGPLv3 allows commercial use, distribution and modification of the sources. The license protects the copyright of the original authors by mandating that *any* derivative work, be it a modification or a different distribution, still be licensed under the terms of the LGPLv3. This point is very important for PCMSOLVER: anyone can use the library without alerting or asking permission from the original authors. However, if modifications, trivial or not, are made, they have to be licensed under the same terms. This makes more likely that such modifications will be submitted back to the main development line for general improvement of the library [66]. *Open-source* and *open data* practices are a heated topic of debate in the computational sciences community [70, 62] and quantum chemistry has had its fair share of lively discussions [53, 79, 73]. There is no private PCMSOLVER development repository. We decided to have the library fully in the open early on in its development. We believe that an open code review process is essential to guarantee scientific reproducibility and this offsets concerns of being scooped by competitors. Krylov et al. [79] noted that open-source at all costs can come with the steep cost of lowered code quality and sloppy maintenance, possibly exacerbating reproducibility issues. However, industry-strength software has and continues to be built by the open-source community. We argue that *more openness* in the computational sciences can have the same transformative effect that it has had in building successful compilers (the GNU compiler collection), operating system kernels (BSD and Linux) and visualization software (ParaView), to just name a few examples. It is our conviction that the gatekeeping model is more detrimental than helpful [73], especially for the modular programming paradigm we advocate. Open-source development has accrued a host of cloud-based services that make advanced maintenance operations trivial to set up and leverage. These include, but are not limited to, continuous integration,[17] static[18] and dynamic[19] code analyses, code coverage evaluation[20] and continuous delivery. The use of Git as distributed version control system (DVCS), together with one of its online front-ends[21] has revolutionized the way open-source software is developed [66]. Public issue tracking and code review have become ubiquitous tools. Both help build better software and are an interactive teaching resource for inexperienced developers joining a new project. All these services and code development techniques can and are used in closed-source development. However, reproducibility, sustainability and extensibility of the software ecosytem in quantum chemistry in particular, and the computational sciences in general, can be more effectively established within an open-source framework. The opportunities for collaboration and the scientific impact will be greater for projects adopting open-source modular development. External contributions, such as improving documentation, reporting bugs, adding new features, are encouraged for the greater benefit of the community at large.

---

[16] Full legal text of the license available from the Free Software Foundation: https://www.gnu.org/licenses/lgpl.html. A condensed version can be found here: https://choosealicense.com/licenses/lgpl-3.0/.

[17] Travis CI: https://travis-ci.org/, AppVeyor CI: https://www.appveyor.com/

[18] Coverity Scan: https://scan.coverity.com/

[19] Valgrind: http://valgrind.org/, AddressSanitizer: https://clang.llvm.org/docs/AddressSanitizer.html, ThreadSanitizer: https://clang.llvm.org/docs/ThreadSanitizer.html

[20] Codecov: https://codecov.io/

We use Git as DVCS[22] for PCMSOLVER and we decided to host the code publicly on GitHub: https://github.com/pcmsolver/pcmsolver. Through Github:

- Users and developers can *open issues* to report bugs, request new features, propose improvements.[23]
- Developers can contribute to the code through pull requests (PRs).[24]
- All code changes are automatically tested using the continuous integration (CI) service Travis.[25] CI guarantees that code changes do not break existing functionality.

GitHub lets developers comment on both issues and PRs so that their relevance can be triaged. The best course of action emerges as a consensus decision. The discussions are complementary to the documentation as a learning resource for experienced and novice developers alike. Figure 7 shows the GitHub user interface for issues and PRs.

We have adopted a fully public fork-and-pull-request (F&PR) workflow, where every proposed changeset has to go through a code review and approval process. A *fork* is a full copy of the canonical repository (https://github.com/PCMSolver/pcmsolver) under a different namespace (https://github.com/Acellera/pcmsolver, for example). The fork is completely independent from the canonical repository and can even diverge from it. The code changes are developed on a *branch* of the *fork*. When completed, the developer submits the changes for review through the web interface: a PR is opened, requesting that the changes from the *source branch* on the fork be merged into a *target branch* in the canonical repository. The PR will include a full `diff` and a brief description and motivation of the proposed changes. Once the PR is open, the new code is automatically tested on Travis. A bot will pre-review the changes based on a set of simple rules. Core developers of PCMSOLVER will then review the contribution and discuss additional changes to be made. Eventually, if all the tests are passing and a developer approves the suggested contribution, the changes are merged into the target branch. The target branch is usually the `master` branch, that is, the main development branch.

A sane versioning scheme is of paramount importance for successful API development [116]. PCMSOLVER uses semantic versioning.[26] Every new release gets a version number of the form `vX.Y.Z-d`:

- `X` is the *major version*. It is only incremented (bumped) when backwards-incompatible changes are introduced. For

---

[21] https://github.com/, https://gitlab.com/

[22] Official documentation for Git can be found here https://git-scm.com/. Git is the *de facto* standard for DVCS, but it can be a daunting task to learn to use it properly. Fortunately, many tutorials are available online. See for example https://coderefinery.github.io/git-intro/ and http://gitimmersion.com/

[23] PCMSOLVER issue tracker: https://github.com/PCMSolver/pcmsolver/issues

[24] PCMSOLVER past and current PRs: https://github.com/PCMSolver/pcmsolver/pulls

[25] PCMSOLVER Travis CI page: https://travis-ci.org/PCMSolver/pcmsolver

[26] https://semver.org/

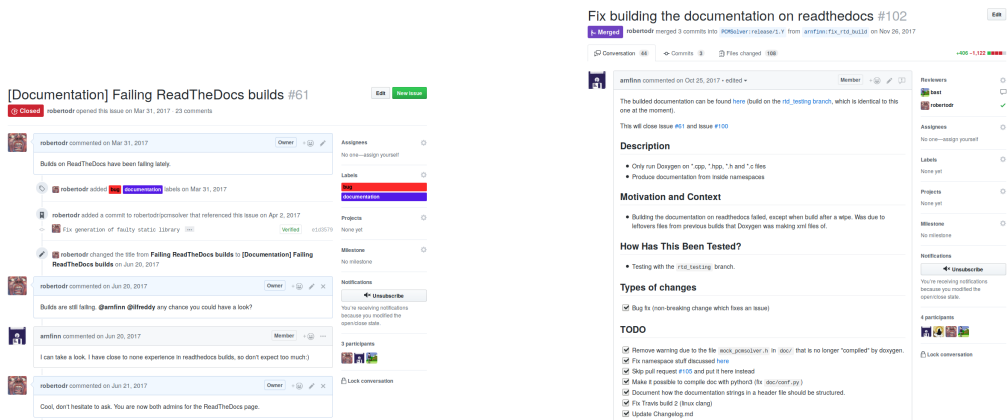

**(a)** A GitHub issue reporting failing documentation builds. **(b)** A GitHub PR with changes to fix the posted issue.

**FIGURE 7** The GitHub user interface for issues and PRs. Both can be extensively discussed and updated until a consensus decision is reached on the best solution for the given problem.

example, developers decided to rename one or more API functions or the parameter packs were changed. These types of changes are rare and are announced timely with deprecation notices.
- `Y` is the *minor version*. It is bumped when new functionalities or non-breaking API changes are introduced.
- `Z` is the *patch version*. Bumping happens whenever a bug is fixed, without adding functionality, nor breaking the API.
- `d` is the *descriptor*. This is an optional component in the version number. It is used to mark unstable (`alpha` or `beta`) or stable but not yet final (release candidates: `rc`) releases.

When enough non-API breaking new functionality accumulates, we prepare a new minor release. This is done by creating a release branch from the `master` branch for a new release, with the format `release/vX.Y`. Such a branch will never be merged back to the `master` branch. It will never receive new features, only bug fixes cherry-picked from the `master` branch. New versions are assigned as Git tags and can be browsed through the GitHub web interface.[27] We keep a detailed change log that serves as a digest of noteworthy changes between versions. We use the GitHub-Zenodo integration to make the project citable and keep track of the citations.[28] Each new release automatically gets a digital object identifier (DOI) from Zenodo. The project can be cited by its global DOI (`10.5281/zenodo.1156166`) that always resolves to the latest released version.

Finally, documentation is written in reStructuredText (`.rst`) format[29] and a webpage can be generated using the Sphinx tool [12]. All code changes applied to the `master` and `release` branches trigger an automatic build of the documentation, which is deployed *via* ReadTheDocs to the website `https://pcmsolver.readthedocs.io` We write code comments in the Doxygen [8] markup language. We use the Doxygen tool to parse the sources and produce documentation for almost all functions and classes in PCMSOLVER. The Breathe [5] plugin to Sphinx integrates the code and end-user documentation.

---

[27] https://github.com/PCMSolver/pcmsolver/releases
[28] https://zenodo.org/
[29] http://docutils.sourceforge.net/rst.html

# 6 | SHOWCASE: PCMSOLVER IN ACTION

The PCMSOLVER library is currently interfaced with the following QC codes, written in a variety of languages:

DIRAC **(Fortran 77)**  A relativistic quantum chemistry program package, implementing, among others, linear and nonlinear response theory and Kramers-restricted correlated methods [1].

DALTON **(Fortran 77)**  A general-purpose program package with emphasis on high-order molecular response properties [2, 16].

LSDALTON **(Fortran 90)**  A linear-scaling program package for the calculation of linear and nonlinear response properties [3, 16].

PSI4 **(C++11, Python)**  An open-source program implementing methods ranging from SCF to CC and multiconfigurational SCF, with a strong emphasis on extensibility and fast method development [141, 100].

RESPECT **(Fortran 90)**  A relativistic DFT quantum chemistry program package, featuring efficient, parallel implementations of, among other methods, real-time time-dependent (TD)-DFT for closed- and open-shell systems [4].

KOALA **(Fortran 90)**  A program package implementing WFT-in-DFT and DFT-in-DFT embedding methods for molecular properties [67, 127].

MADNESS **(C++11)**  A massively parallel implementation of multiresolution analysis (MRA) for chemistry and physics applications [58].

Many of these codes did not have PCM capabilities before the interface was put in place and, in the case of KOALA and MADNESS, the program developers required little to no assistance from the PCMSOLVER developers to implement the coupling with the library. Our license plays well with closed-source software: not all of the codes listed are open-source, while some of them (DALTON and LSDALTON) only recently switched to the LGPLv2.1 and an open collaboration workflow.[30] In all but one of the above mentioned cases, the implementation of the interface resulted in a collaboration between the authors and the host code developers. This further proves our point that the availability of open modules with well-defined interfaces is one of the keys to enhancing collaboration and dissemination of ideas in quantum chemistry.

We described the theory for and the implementation of the interface with the DIRAC program in 2015 [34]. This work proved the principle that a well-defined API for the PCM could be formulated abstractly from the details of the quantum chemical method of choice. Our implementation in DIRAC also showed the importance of optimizing the calculation of the MEP one-electron integrals for large grids of points. Being agnostic of its use for PCM calculations, our efficient implementation of such integrals for 4-component wave functions in DIRAC is currently also used to export the MEP on the grid used in frozen-density embedding (FDE) calculations [68, 97]. Our work paved the way for the recent implementation of a relativistic SCF method including polarizable embedding (PE) terms, published by Hedegård et al. [63].

Building on our experience with DIRAC, we introduced the PCMSOLVER library into the structurally similar DALTON and LSDALTON codes. In contrast to DALTON, which already had a PCM implementation available, the interface in LSDALTON provided new functionality to the users. The LSDALTON interface was then used to assess the accuracy attainable in quantum chemical calculations with PCM in conjunction with a wavelet Galerkin solver. This work was a collaboration with mathematicians and LSDALTON developers and was described by Bugeanu et al. [23]. Figures 8a and 8b summarize our findings: the wavelet Galerkin solver can attain much higher accuracy, at the expense of introducing a much larger grid of points on the cavity surface. Approximations in the integral evaluation subroutines will have to be introduced, an area that we are currently investigating.

---

[30]The DALTON and LSDALTON public Git repositories are hosted on GitLab: `https://gitlab.com/dalton`

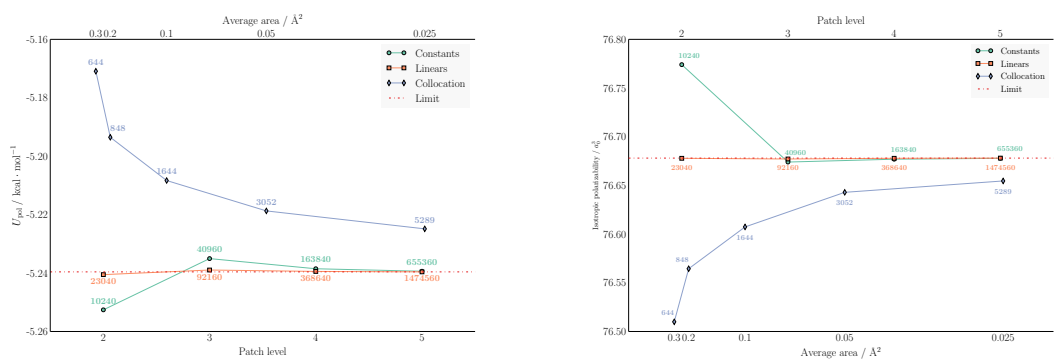

**(a)** Convergence of $U_{\mathrm{pol}}$ with the number of MEP evaluation points on the cavity surface. Lower axis: patch level in the wavelet Galerkin discretization. Upper axis: average area for the collocation tesselation.

**(b)** Convergence of $\alpha_{\mathrm{iso}}$ with the number of MEP evaluation points on the cavity surface. Lower axis: average area for the collocation tesselation. Upper axis: patch level in the wavelet Galerkin discretization.

**FIGURE 8** Convergence of $U_{\mathrm{pol}}$ and $\alpha_{\mathrm{iso}}$ for benzene with respect to the number of MEP evaluation points on the cavity surface, when using collocation, piecewise constant and piecewise linear wavelet Galerkin solvers. The number of such points is reported as an annotation of the data points. All Hartree–Fock (HF)/6-31G calculations performed with LSDALTON. Figures reproduced from Bugeanu et al. [23] - Published by the PCCP Owner Societies.

Our work in DALTON concentrated on the calculation of high-order molecular properties, motivated by the open-ended implementation of response theory that has been ongoing in our group [139, 120, 50]. As briefly discussed in section 2.2, the variational formalism greatly simplifies formal derivations, a fact that we found especially true for response theory. The extended quantum/classical polarizable quasienergy Lagrangian formalism, allows us to leverage the recursive implementation of Ringholm et al. [120]. The formalism and its implementation were applied to the calculation of one- to five-photon absorption strengths of small chromophores in different solvents [35]. Figure 9 shows our result for *para*-dinitrobenzene, a centrosymmetric molecule. Using a nonequilibrium response formulation for the solvent results in discontinuities in the enhancement as a function of solvent polarity.

Two new extensions to the PCMSOLVER library will be released to the public soon. One such extension is a real-time propagation scheme for the solvent effect both with the equilibrium abd the delayed schemes described by Corni et al. [29] and Ding et al. [41]. In the former, the polarization immediately responds to changes in the solute density; in the latter, a retardation effect is introduced due to the solvent permittivity being nonlocal in time. To illustrate this development, Figure 10 shows preliminary results for the one-photon absorption spectra of the uranyl ion $UO_2^{2+}$ with a 4-component relativistic Hamiltonian.. This is part of an ongoing collaboration with the RESPECT developers. We coupled the efficient and parallelized real-time propagation algorithm with the PCM [38]. Our implementation can tackle the rather large systems arising when heavy-element containing systems are of interest. Coupling with the PCM introduces a negligible overhead in the real-time propagation and this method will surely help shed light into the interplay of relativistic and solvent effects. This new functionality has not yet been released, but a version of RESPECT including the interface with PCMSOLVER is already available for the calculation of SCF energies and first-order electric and magnetic properties of closed- and open-shell systems [39].

The other extension is the implementation of the FQ classical polarizable model within PCMSOLVER. Section 2 showed the striking similarities of continuum and explicit classical polarizable models for the environment. The FQ model is straightforward to implement on top of the PCM infrastructure we have put together, since its input and output with the QC host program are *identical* to those for the PCM. Hence, *any code* currently interfaced with PCMSOLVER can have access to our FQ implementation by simply *upgrading* their version of the library and preparing appropriate input files. There is no additional coding involved. Figure 11 shows our preliminary results for the one-photon absorption spectrum of the rhodamine 6G chromophore, a promising dye for nonlinear photonic applications [96, 95]. The calculations were performed using a development version of LSDALTON. When released it will further enrich the set of methods available to users of this code.

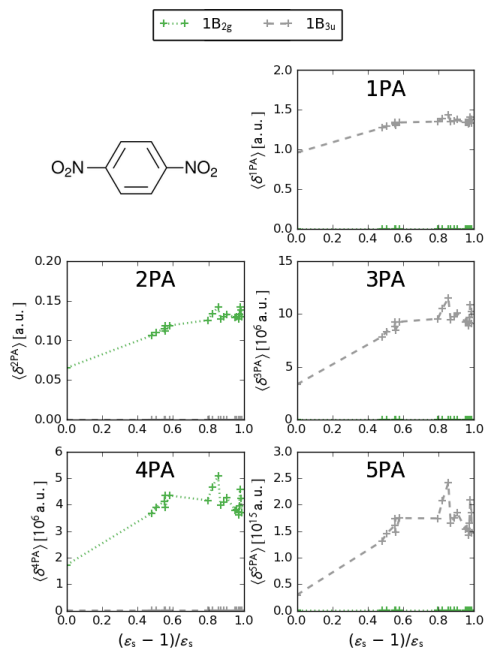

**FIGURE 9** One- to five-photon absorption strengths ($\langle\delta^{\mathrm{MPA}}\rangle$) in atomic units for the centrosymmetric molecule *para*-dinitrobenzene. The data is plotted for two selected electronic excitations and as a function of increasing solvent polarity $\frac{\varepsilon_s - 1}{\varepsilon_s}$, where $\varepsilon_s$ is the static permittivity. All CAM-B3LYP/aug-cc-pVDZ response calculations were performed using DALTON and the nonequilibrium formulation for the solvent terms. Figures reproduced from Di Remigio et al. [35] - Published by the PCCP Owner Societies.

# 7 | THE FUTURE: LESSONS LEARNT, THE ROAD AHEAD AND SOME QUESTIONS

From day one, the informal motto of PCMSOLVER has been *Plug the solvent in your favorite QM code*. We have built the library striving for:

- QM host program agnosticism.
- Intuitive API for QM host program developers.
- Open and inclusive code development workflow.
- Extensible internal code structure.

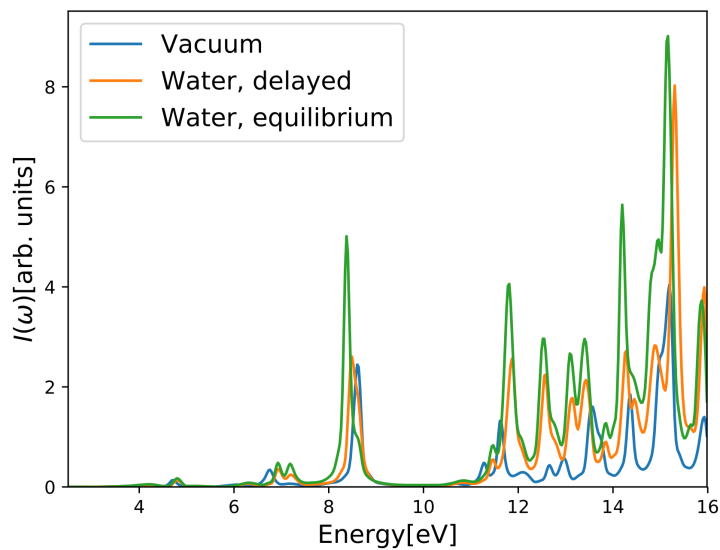

**FIGURE 10** 4-component one-photon absorption spectra of the uranyl ion $UO_2^{2+}$ in vacuum and in water. The spectra were obtained from a real-time TD-DFT simulation using the PBE functional and a triple-zeta quality basis set, ( [33$s$29$p$20$d$13$f$4$g$2$h$] for U, [11$s$6$p$3$d$2$f$] for O), was employed, together with the resolution-of-the-identity (RI)-J algorithm (fitting bases: [41$s$37$p$37$d$24$f$24$g$15$h$] for U, [14$s$8$p$8$d$4$f$4$g$3$h$] for O). Calculations were ran using a development version of RESPECT [118, 78, 4, 38].

We have achieved the former two points. The reader does not have to take our word for it, though. PCMSOLVER is interfaced with many QM host programs, enlarging the breadth of applications these programs can tackle. The development of the library is not, however, just a success story. Achieving the latter two points has proved much more challenging.

To move forward on the road ahead we will have to attract contributions from more developers, improve our API, introduce new features and interface with more QC codes. Some new features, such as an implementation of the FQ polarizable force field and of the real-time evolution proposed by Corni et al. [29] are almost ready for release.

The variational formulation of the PCM [86] is a convenient tool for deriving the quantum/classical coupling terms in theories as diverse as SCF [87], CC [36, 33] and arbitrary-order response theory [35]. From a theoretical perspective, it provides a much cleaner route to the derivation of the working equations. Recasting the coupled problem as a variational minimization also gives insight into alternative algorithmic realizations. But the advantages of the approach are greater still. As shown by Lipparini, *explicit* classical polarizable models admit a variational formulation [82, 83, 89, 85]. Three-layer coupling, as realized, for example, in QM/MM/Continuum protocols, are then trivial to derive. The fundamental similarity between classical polarizable models has been recognized long before the advent of the, clearly superior, variational formulation. However, a similarly unified implementation of these models has yet to appear. As the formulation of such diverse models can be put on an equal footing, the same must also be true for their computational implementation. The question is how can such a task be accomplished. PCMSOLVER offers a starting point. There is a discussion issue open on GitHub: we hope many in the community will join us in this effort.[31]

---

[31] https://github.com/PCMSolver/pcmsolver/issues/139

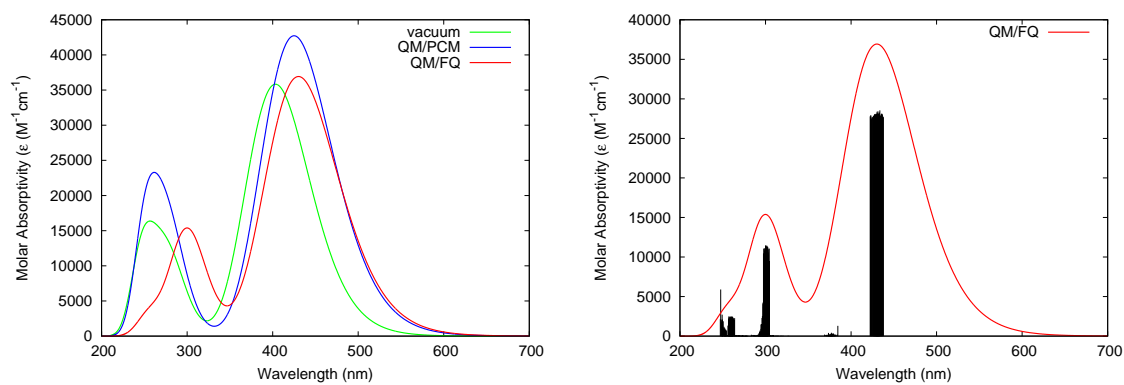

**(a)** One-photon absorption spectra of rhodamine 6G in water (PCM and FQ) and vacuum. The FQ spectrum is an average over 100 snapshots.

**(b)** One-photon absorption spectrum of rhodamine 6G in water using FQ. The stick and average (over 100 snapshots) spectra are also shown.

**FIGURE 11** CAM-B3LYP/6-31+G* one-photon absorption spectra of rhodamine 6G in vacuum and water (PCM and FQ). Calculations were ran using density fitting (fitting basis: df-def2) with a development version of LSDALTON [16, 3]. A solvation shell of 20 Å was used in the FQ calculations. The spectra are convoluted with Gaussian lineshapes. Figures reproduced courtesy of Tommaso Giovannini (Scuola Normale Superiore, Pisa).

**ACKNOWLEDGEMENTS**

The authors would like to acknowledge the users and programmers that have helped in the development of the PCMSOLVER library through their bug reporting, enhancement suggestions and code patches.

RDR is particularly indebted with Lori A. Burns (Georgia Tech), for suggestions on CMake and code infrastructure, Radovan Bast (University of Tromsø - The Arctic University of Norway), for discussions on open-source software, Filippo Lipparini (Università di Pisa) for many clarifications on the variational formulation of QM/classical polarizable models, and Andrew M. James (Virginia Tech) for providing comments on an early version of the manuscript.

The authors acknowledge partial support by the Research Council of Norway through its Centres of Excellence scheme, project number 262695. RDR also acknowledges support by the Research Council of Norway through its Mobility Grant scheme, project number 261873.

The authors received support by the Norwegian Supercomputer Program through a grant for computer time (Grant No. NN4654K). RDR acknowledges the Advanced Research Computing Center at Virginia Tech for providing the necessary computational resources and technical support for some of the calculations reported here.

ROBERTO DI REMIGIO ET AL. | 39[80] Lange AW, Herbert JM. A smooth, nonsingular, and faithful discretization scheme for polarizable continuum models: the switching/Gaussian approach. J Chem Phys 2010 Dec;133(24):244111. http://dx.doi.org/10.1063/1.3511297.

[81] Lange AW, Herbert JM. Polarizable Continuum Reaction-Field Solvation Models Affording Smooth Potential Energy Surfaces. J Phys Chem Lett 2010;1(2):556–561. http://dx.doi.org/10.1021/jz900282c.

[82] Lipparini F, Barone V. Polarizable Force Fields and Polarizable Continuum Model: A Fluctuating Charges/PCM Approach. 1. Theory and Implementation. J Chem Theory Comput 2011 Nov;7(11):3711–3724. http://pubs.acs.org/doi/abs/10.1021/ct200376z.

[83] Lipparini F, Cappelli C, Barone V. Linear Response Theory and Electronic Transition Energies for a Fully Polarizable QM/Classical Hamiltonian. J Chem Theory Comput 2012 Nov;8(11):4153–4165. http://dx.doi.org/10.1021/ct3005062.

[84] Lipparini F, Cappelli C, Barone V. A gauge invariant multiscale approach to magnetic spectroscopies in condensed phase: General three-layer model, computational implementation and pilot applications. J Chem Phys 2013;138(23):234108. http://link.aip.org/link/JCPSA6/v138/i23/p234108/s1&Agg=doi.

[85] Lipparini F, Mennucci B. Perspective: Polarizable continuum models for quantum-mechanical descriptions. J Chem Phys 2016 Apr;144(16):160901. http://scitation.aip.org/content/aip/journal/jcp/144/16/10.1063/1.4947236?TRACK=RSS.

[86] Lipparini F, Scalmani G, Mennucci B, Cancès E, Caricato M, Frisch MJ. A variational formulation of the polarizable continuum model. J Chem Phys 2010 Jul;133(1):014106. http://dx.doi.org/10.1063/1.3454683.

[87] Lipparini F, Scalmani G, Mennucci B, Frisch MJ. Self-Consistent Field and Polarizable Continuum Model: A New Strategy of Solution for the Coupled Equations. J Chem Theory Comput 2011 Mar;7(3):610–617. http://dx.doi.org/10.1021/ct1005906.

[88] Lippman SB, Lajoie J, Moo BE. C++ Primer. 5 edition ed. Addison-Wesley Professional; 2012. https://www.amazon.com/Primer-5th-Stanley-B-Lippman/dp/0321714113/ref=sr_1_1?ie=UTF8&qid=1522430097&sr=8-1&keywords=lippman+c%2B%2B+primer.

[89] Loco D, Polack É, Caprasecca S, Lagardère L, Lipparini F, Piquemal JP, et al. A QM/MM Approach Using the AMOEBA Polarizable Embedding: From Ground State Energies to Electronic Excitations. J Chem Theory Comput 2016 Aug;12(8):3654–3661. http://dx.doi.org/10.1021/acs.jctc.6b00385.

[90] Mantina M, Chamberlin AC, Valero R, Cramer CJ, Truhlar DG. Consistent van der Waals Radii for the Whole Main Group. J Phys Chem A 2009 14 May;113(19):5806–5812.

[91] Mennucci B, Cammi R. Continuum Solvation Models in Chemical Physics: From Theory to Applications. Wiley; 2008. https://books.google.no/books?id=6Om2gDR41rwC.

[92] Merali Z. Computational Science: ...Error. Nature 2010 14 Oct;467(7317):775–777.

[93] Messina R. Image charges in spherical geometry: Application to colloidal systems. J Chem Phys 2002 Dec;117(24):11062–11074. https://doi.org/10.1063/1.1521935.

[94] Miertuš S, Scrocco E, Tomasi J. Electrostatic interaction of a solute with a continuum. A direct utilizaion of ab initio molecular potentials for the prevision of solvent effects. Chem Phys 1981 Feb;55(1):117-129. http://linkinghub.elsevier.com/retrieve/pii/0301010481850902.

[95] Milojevich CB, Silverstein DW, Jensen L, Camden JP. Surface-Enhanced Hyper-Raman Scattering Elucidates the Two-Photon Absorption Spectrum of Rhodamine 6G. J Phys Chem C 2013 Feb;117(6):3046–3054. http://dx.doi.org/10.1021/jp3094098.